\pdfoutput=1
\documentclass[11pt,twoside,a4paper,cmspaper,final,collab]{cms-tdr}
\def\svnVersion{e4fe1d8}\def\svnDate{2025/05/27}\def\cmsCernNoTag{CERN-EP-2024-210}\def\cmsCernDate{\today}\def\cmsMessage{Published in Physical Review D as \href{http://dx.doi.org/10.1103/PhysRevD.111.092014}{\doi{10.1103/PhysRevD.111.092014}.}}
\begin{document}\cmsNoteHeader{HIG-21-019}

\newcommand{\Dkinbkg}{\ensuremath{\mathcal{D}^\text{kin}_\text{bkg}}\xspace}
\newcommand{\RelMassErr}{\ensuremath{\delta m_{4\Pell}/m_{4\Pell}}\xspace}
\newcommand{\mZone}{\ensuremath{m(\PZ_{1})}\xspace}
\newcommand{\mZ}{\ensuremath{{m}_{\PZ}}\xspace}
\newcommand{\ggZZ}{\ensuremath{\Pg\Pg \to \PZ\PZ}\xspace}
\newcommand{\Hfourl}{\ensuremath{\PH \to 4\Pell}\xspace}
\newcommand{\Eq}[1]{Eq.~(\ref{#1})}
\newcommand{\Eqs}[3]{Eqs.~(\ref{#1}) #3 (\ref{#2})}
\newcommand{\ROOT}{\textsc{root}\xspace}
\newcommand{\jhugen}{\textsc{JHUGen}\xspace}
\newcommand{\minlo}{\textsc{MINLO}\xspace}
\newcommand{\mela}{\textsc{mela}\xspace}
\newcommand{\hnnlo}{\textsc{hnnlo}\xspace}
\newcommand{\F}{\ensuremath{\cmsSymbolFace{F}}\xspace}
\newcommand{\WW}{\ensuremath{\PW\PW}\xspace}
\newcommand{\ZZ}{\ensuremath{\PZ\PZ}\xspace}
\newcommand{\VVV}{\ensuremath{\cmsSymbolFace{VVV}}\xspace}
\newcommand{\VBF}{\ensuremath{\mathrm{VBF}}\xspace}
\newcommand{\VH}{\ensuremath{\PV\PH}\xspace}
\newcommand{\WH}{\ensuremath{\PW\PH}\xspace}
\newcommand{\ZH}{\ensuremath{\PZ\PH}\xspace}
\newcommand{\tqH}{\ensuremath{\PQt\PQq\PH}\xspace}
\newcommand{\ttH}{\ensuremath{\PQt\PAQt\PH}\xspace}
\newcommand{\bbH}{\ensuremath{\PQb\PAQb\PH}\xspace}
\newcommand{\ggH}{\ensuremath{\Pg\Pg\PH}\xspace}
\newcommand{\ZX}{\ensuremath{\PZ+\PX}\xspace}
\newcommand{\Hboson}{Higgs boson\xspace}
\newcommand{\Hell}{\ensuremath{\PH\to4\Pell}\xspace}
\newcommand{\GH}{\ensuremath{\Gamma_\PH}\xspace}
\newcommand{\Gref}{\ensuremath{\Gamma_0}\xspace}
\newcommand{\GHSM}{\ensuremath{\Gamma_\PH^{\mathrm{SM}}}\xspace}
\newcommand{\mH}{\ensuremath{m_{\PH}}\xspace}
\newcommand{\mell}{\ensuremath{m_{4\Pell}}\xspace}
\newcommand{\bbar}{\ensuremath{\PQb\PAQb}\xspace}
\newcommand{\muF}{\ensuremath{\mu_{\F}}\xspace}
\newcommand{\muV}{\ensuremath{\mu_{\PV}}\xspace}
\newcommand{\Dbkg}{\ensuremath{{\mathcal{D}}_{\text{bkg}}}\xspace}
\newcommand{\Dint}{\ensuremath{{\mathcal{D}}_{\text{int}}}\xspace}
\newcommand{\Dbsi}{\ensuremath{{\mathcal{D}}_{\text{bsi}}}\xspace}
\newcommand{\Dbkgkin}{\ensuremath{{\mathcal{D}}^{\text{kin}}_{\text{bkg}}}\xspace}
\newcommand{\offshell}{off-shell\xspace}
\newcommand{\onshell}{on-shell\xspace}
\newcommand{\HVV}{\ensuremath{\PH\PV\PV}\xspace}
\ifthenelse{\boolean{cms@external}}{\providecommand{\cmsTable}[1]{#1}}{\providecommand{\cmsTable}[1]{\resizebox{\textwidth}{!}{#1}}}

\providecommand{\cmsTable}[1]{\resizebox{\textwidth}{!}{#1}}

\cmsNoteHeader{HIG-21-019}
\title{Measurement of the Higgs boson mass and width using the four-lepton final state in proton-proton collisions at \texorpdfstring{$\sqrt{s} = 13\TeV$}{sqrt(s) = 13 TeV}}

\author*[cern]{A. Cern Person}

\date{\today}

\abstract{A measurement of the Higgs boson mass and width via its decay to two \PZ bosons is presented. Proton-proton collision data collected  by the CMS experiment, corresponding to an integrated luminosity of 138\fbinv at a center-of-mass energy of 13\TeV, is used.  The invariant mass distribution of four leptons in the on-shell Higgs boson decay is used to measure its mass and constrain its width.  This yields the most precise single measurement of the Higgs boson mass to date, $125.04 \pm 0.12\GeV$,  and an upper limit on the width $\Gamma_{\PH} < 330\MeV$ at 95\% confidence level.  A combination of the on- and off-shell Higgs boson production decaying to four leptons is used to determine the Higgs boson width, assuming that no new virtual particles affect the production,  a premise that is tested by adding new heavy particles in the gluon fusion loop model. This result is combined with a previous CMS analysis of the off-shell Higgs boson production with decay to two leptons and two neutrinos,  giving a measured Higgs boson width of $3.0^{+2.0}_{-1.5}\MeV$, in agreement with the standard model prediction of 4.1\MeV.  The strength of the off-shell Higgs boson production is also reported.  The scenario of no off-shell Higgs boson production is excluded at a confidence level corresponding to 3.8 standard deviations.}

\hypersetup{
pdfauthor={CMS Collaboration},
pdftitle={Measurement of the Higgs boson mass and width using the four-lepton final state in proton-proton collisions at sqrt(s) = 13 TeV},
pdfsubject={CMS},
pdfkeywords={CMS, Higgs boson, mass, width, four-lepton, 4l}}

\maketitle 

\section{Introduction}
\label{sec:Introduction}
The standard model (SM) of particle physics postulates the existence of a Higgs field responsible for the
generation of the masses of fundamental particles. The excitation of this
field is known as the Higgs boson ($\PH$)~\cite{StandardModel67_1, Englert:1964et,Higgs:1964ia,
Higgs:1964pj,Guralnik:1964eu,StandardModel67_2,StandardModel67_3}.
The properties of the \Hboson, observed with a mass of approximately 125\GeV by the ATLAS and CMS
Collaborations~\cite{Aad:2012tfa,Chatrchyan:2012xdj,Chatrchyan:2013lba} at the CERN LHC, are found to be consistent with
the expectations of the SM~\cite{ATLASnature,CMSnature}. The mass of the \Hboson (\mH) is a free parameter of the model and, 
since it determines all other Higgs properties, should be measured with as high precision as possible.
For example, the Higgs boson couplings to vector bosons strongly depend on the \Hboson mass and are precisely predicted by the SM.
Another important \Hboson characteristic is its lifetime, predicted by the SM to be $1.6\times10^{-22}$\,s, corresponding to a total width (\GH) of $4.1$\MeV~\cite{deFlorian:2016spz},
as predicted precisely within the SM for $\mH = 125 \GeV$.
A deviation from the SM prediction would point to either anomalous \Hboson couplings or its decay to yet undiscovered particles.

The ATLAS and CMS Collaborations measured the \Hboson mass to be $125.09 \pm 0.24\GeV$ \cite{Aad:2015zhl} using $\sqrt{s}=7$ and 8\TeV proton-proton ($\Pp\Pp$) collision data from the 2011--2012 data-taking periods (Run 1), corresponding to a total integrated luminosity per experiment of 25\fbinv.
This result has been superseded by both collaborations.
The ATLAS experiment measured the \Hboson mass to be $125.11 \pm 0.11 (\pm 0.09)\GeV$ \cite{ATLAS_mass}, combining the $\PH \to \PGg\PGg$ and \Hfourl ($\Pell = \Pe$, $\PGm$) channels from Run 1 and data collected at $\sqrt{s}=13\TeV$ in 2015--2018 (Run~2).
The value in parentheses is the statistical uncertainty only.
The most recent CMS result, also using the $\PH \to \PGg\PGg$ and \Hfourl channels and including Run 1 and 36\fbinv of $\sqrt{s}=13\TeV$ data from 2016, is 
\mH = $125.38 \pm 0.14$ $(\pm 0.11) \GeV$.
Measurements from ATLAS and CMS using only the \Hfourl channel and 2016 data are $124.94 \pm 0.18 (\pm 0.17)\GeV$ and $125.26 \pm 0.21 (\pm 0.19)\GeV$, respectively.

Considering only on-shell Higgs boson production, CMS set an upper limit 
on the \Hboson width $\GH < 1.10\GeV$ at 95\% confidence level (\CL), 
limited by the four-lepton invariant mass resolution~\cite{Khachatryan:2014jba,Sirunyan:2017exp}.
Both the ATLAS and CMS experiments have also set limits on 
$\GH$~\cite{Khachatryan:2014iha,Aad:2015xua,Khachatryan:2015mma,
Khachatryan:2016ctc,Aaboud:2018puo,Sirunyan:2019twz,CMS:2022ley}
from an \offshell production method~\cite{Caola:2013yja,Kauer:2012hd,Campbell:2013una},
which relies on the measurement of the ratio of off- to on-shell production rates.
Considering both gluon fusion (\ggH) and electroweak (EW) processes, the most recent measurements are
$\GH= 3.2^{+2.4}_{-1.7}\MeV$ \cite{CMS:2022ley} and $\GH = 4.3^{+3.3}_{-2.5}\MeV$~\cite{atlascollaboration2023evidence} by CMS and ATLAS, respectively.
Finally, from an upper limit on the \Hboson flight distance in the detector,
CMS set a lower limit of $\GH > 3.5\times10^{-9}$\MeV at 95\% \CL~\cite{Khachatryan:2015mma}.

This paper reports an updated CMS measurement of the \Hboson mass and width using \onshell production and the \Hfourl decay.
The data sample includes 138\fbinv of $\Pp\Pp$ collision data at $\sqrt{s}=13\TeV$ collected in 2016--2018, in combination with the Run 1 data.
Compared to the previous CMS \onshell \Hboson measurement in this channel~\cite{Sirunyan:2017exp}, 
the statistical and systematic uncertainties affecting \mH have been reduced by including the beam spot in a refit of the muon tracks; 
adopting an improved event categorization procedure; and performing a detailed study of the lepton momentum scale and resolution.

A measurement of the relative off- and on-shell \Hboson production offers direct information about 
\GH~\cite{Caola:2013yja,Kauer:2012hd,Campbell:2013una}.
For each \Hboson production mechanism $j$, with subsequent decay to four leptons, 
the on- and \offshell cross sections $\sigma_j$ are proportional to
\begin{equation}
\label{eq:resonant}
\sigma_j^\text{\onshell} \propto \frac{g_p^2 g_d^2}{\GH}  \propto \mu_j^\text{\onshell} 
\quad\text{and}\quad
\sigma_j^\text{\offshell} \propto {g_p^2 g_d^2} \propto \mu_j^\text{\onshell}  \ \GH,
\end{equation}
where $g_p$ and $g_d$ are the couplings associated with the \Hboson production and decay, respectively, and
$\mu_j^\text{\onshell}$ is the \onshell signal strength, defined as the ratio of the observed number 
of \onshell four-lepton events relative to the SM expectation.
The \onshell signal strength inherently includes the dependence on the \Hboson couplings and \GH.
The signal strength is denoted as \muF for \Hboson production mechanisms driven by fermion couplings, 
\ie, production via \ggH or in association with a \ttbar (\ttH) or \bbar pair (\bbH). 
For EW production, \ie, production via vector boson fusion (\VBF) 
or in association with a $\PW$ or $\PZ$ boson (\VH), the ratio is denoted as \muV.
Contrary to gluon fusion and EW \onshell production, there is sizable destructive interference between 
the \Hboson signal and the nonresonant four-lepton production in the \offshell region~\cite{Lee:1977yc,Kauer:2012hd}.
This interference is crucial for maintaining unitarity and scales with $\sqrt{ \smash[b]{\mu_j^\text{\onshell} \GH}}$.
 
In the described technique for measuring \GH, it is anticipated that the ratio of the couplings governing off- and \onshell 
production matches the SM prediction. 
In particular, it is assumed that the dominant production mechanism is $\ggH$ rather than quark-antiquark annihilation.
The dominance of the $\ggH$ production mechanism has been thoroughly tested in the on-shell regime~\cite{deFlorian:2016spz, CMSnature}. 
It is also assumed that beyond-SM particles do not make significant contributions to the $\ggH$ loop within the mass range considered by the analysis.
In this paper, we explicitly test this assumption for the first time through a joint off- and \onshell analysis and find that the 
\GH constraints are not substantially altered.
In our previous \offshell analyses~\cite{Sirunyan:2019twz,CMS:2022ley}, we evaluated the anomalous contributions 
to the $\PH\PV\PV$ vertex (where $\PV$ denotes a \PW boson, \PZ boson, or $\PGg$) in both EW production and \Hboson decay. 
We found that these potential contributions did not significantly affect the \GH bounds.
It is also assumed that no beyond-SM particles, such as higher-mass resonances, significantly contribute within the mass range 
investigated by the analysis. However, such resonances would typically increase the yield of events at higher masses, 
which is not supported by our measurement, and no such resonances have been found in a direct search~\cite{Sirunyan:2018qlb}.
These tests do not address every possible scenario that could impact the measurement of the width, but
a violation of any of the above assumptions would, by itself, indicate the presence of physics beyond the SM.

The \Hboson width may deviate from the SM expectation of 4.1\MeV~\cite{deFlorian:2016spz}
if the \Hboson has non-SM decay channels, or if the known decay modes have non-SM rates. 
Therefore, the direct measurement of the \Hboson width complements searches for \Hboson decays to invisible 
or undetected particles and measurements of the \Hboson couplings to the known SM particles.
For example, if the Higgs boson decays into a pair of unknown particles, potentially candidates 
for dark matter, this would increase the predicted Higgs boson width but 
would not introduce a bias into the measurement techniques used in the current analysis. 

\section{The CMS detector}
\label{sec:Detector}
The central feature of the CMS apparatus is a superconducting solenoid of 6\unit{m} internal diameter, providing a magnetic field of 3.8\unit{T}. Within the solenoid volume are a silicon pixel and strip tracker, a lead tungstate crystal electromagnetic calorimeter (ECAL), and a brass and scintillator hadron calorimeter (HCAL), each composed of a barrel and two endcap sections. Forward calorimeters extend the pseudorapidity coverage provided by the barrel and endcap detectors. Muons are reconstructed in gas-ionization detectors embedded in the steel flux-return yoke outside the solenoid. More detailed descriptions of the CMS detector, together with a definition of the coordinate system used and the relevant kinematic variables, can be found in Refs.~\cite{CMS:2008xjf,CMS:2023gfb}. 

Events of interest are selected using a two-tiered trigger system. The first level, composed of custom hardware 
processors, uses information from the calorimeters and muon detectors to select events at a rate of around 100\unit{kHz}
within a fixed latency of about 4\mus~\cite{Sirunyan:2020zal}. The second level, known as the high-level trigger,
consists of a farm of processors running a version of the full event reconstruction software optimized for fast processing,
and reduces the event rate to around 1\unit{kHz} before data storage~\cite{Hayrapetyan_2024}.

The primary vertex is taken to be the vertex corresponding to the hardest scattering in the event, evaluated using tracking information alone, as described in Section 9.4.1 of Ref.~\cite{CMS-TDR-15-02}.

The electron momentum is estimated by combining the energy measurement in the ECAL with the momentum measurement in the silicon tracker. The transverse momentum (\pt) resolution for electrons with $\pt \approx 45\GeV$ from $\PZ \to \Pe \Pe$ decays ranges from 1.6 to 5\%. It is generally better in the barrel region than in the endcaps, and also depends on the bremsstrahlung energy emitted by an electron as it traverses the material in front of the ECAL~\cite{eleReco,ScaleSmear2}. 

Muons are measured in the pseudorapidity range $\abs{\eta} < 2.4$, with detection planes made using three technologies: drift tubes, cathode strip chambers, and resistive plate chambers. The single-muon trigger efficiency exceeds 90\% over the full $\eta$ range, and the efficiency to reconstruct and identify muons is greater than 96\%. Matching muons to tracks measured in the silicon tracker results in a relative transverse momentum resolution, for muons with \pt up to 100\GeV, of 1\% in the barrel and 3\% in the endcaps. The \pt resolution in the barrel is better than 7\% for muons with \pt up to 1\TeV~\cite{MuReco}.

\section{Data and simulated samples}
\label{sec:Bkg}

The data sample used in this analysis corresponds to integrated luminosities of 36.3, 41.5, and 59.8\fbinv
collected in 2016, 2017, and 2018, respectively,  for a total of 138\fbinv at a center-of-mass energy of 13\TeV.
Events are selected online using the same set of triggers as adopted in Ref.~\cite{Sirunyan:2021rug}. 
They require the presence of at least one lepton (either muon or electron),
or up to three leptons with relaxed \pt conditions. The trigger efficiency relative to the offline selection 
is found to be larger than 99\%, measured in data using $4\Pell$ events collected by the single-lepton triggers. 
It agrees with the expectation from simulation at permille precision. 
Monte Carlo simulations are used to model signal processes, which involve the \Hboson
and background processes.

On-shell \Hboson production through \ggH is simulated using the 
\POWHEG~2.0~\cite{Frixione:2007vw,Bagnaschi:2011tu,Nason:2009ai,Luisoni:2013kna,Hartanto:2015uka,PowhegBox_1,PowhegBox_2}   
event generator at next-to-leading order (NLO) in quantum chromodynamics (QCD). 
Simulation with the \minlo~\cite{Hamilton:2012np} program at NLO in QCD is used for the evaluation of 
systematic uncertainties related to the modeling of jets produced in association with the \Hboson. 
Modeling of associated jet activity in gluon fusion simulation is important for categorization of events across 
a wide range of four-lepton invariant masses for the \Hboson \offshell analysis.

On-shell \Hboson production through \VBF, in association with a $\PW$ or $\PZ$ boson, or with a \ttbar pair,
is simulated using both \POWHEG~2.0 at NLO in QCD and
\jhugen~7.3.0~\cite{Gao:2010qx,Bolognesi:2012mm,Anderson:2013afp,Gritsan:2016hjl,Gritsan:2020pib}
at leading order (LO) in QCD.
Production in association with a \bbar pair or a single top quark is simulated only at LO in QCD with \jhugen.
In the \VBF, \VH, and \ttH production modes, the \jhugen and \POWHEG simulations are explicitly compared
after parton showering, and no significant differences are found in the kinematic observables.
Both on- and off-shell \Hboson production are simulated following 
Refs.~\cite{Sirunyan:2019twz,CMS:2022ley}. 
Therefore, the \jhugen simulation is adopted to describe kinematic distributions in the \VBF, \VH, \ttH, \bbH, and \tqH production
mechanisms in the \onshell region, with the expected yields taken from the higher-order \POWHEG simulation. 
The considered \VH process does not include $\Pg\Pg\to\PZ\PH$ production, which is expected 
to contribute only about 5\% of the \VH cross section~\cite{deFlorian:2016spz}, and is therefore neglected in this analysis. 

The modeling of \offshell \Hboson production incorporates interference effects between diagrams that include the \Hboson 
propagator and those that do not. 
The primary process is gluon fusion $\Pg\Pg \to (\PH^*) \to \ZZ$ / $\PZ\PGg^*\to4\Pell$.
The other process is EW production, which includes vector boson scattering and triple-gauge-boson (\VVV) production.
Both processes are simulated using 
\MCFM~7.0.1~\cite{MCFM,Campbell:2011bn,Campbell:2013una,Campbell:2015vwa},
which is integrated into the \jhugen generator framework.
The \jhugen framework enables the incorporation of anomalous interactions involving 
the \Hboson or gauge bosons in both \ggH and EW production processes. 
The \offshell \Hboson simulation inherently includes \onshell production as well, which is removed 
based on truth-level information from MC simulation and replaced with the dedicated \onshell simulation described above.
The simulation of the EW process is validated by comparing its results to those obtained 
with the \textsc{phantom}~1.3 generator~\cite{Ballestrero:2007}. 
All \offshell production simulations are conducted at LO in QCD, with higher-order corrections 
included through a $K$ factor, as described later.

The \mela~\cite{Gao:2010qx,Bolognesi:2012mm,Anderson:2013afp,Gritsan:2016hjl,Gritsan:2020pib} 
package contains a library of matrix elements from \jhugen for the simulation of the signal channel
and from \MCFM for the background, which enables further reweighting
of the generated \offshell samples, as discussed in the following. 
The \POWHEG program is used to simulate wide resonances with masses ranging from
115\GeV to 3\TeV produced in \ggH, \VBF, or \VH,
and the \jhugen program simulates their decay to four leptons. 
The events from the $\POWHEG + \jhugen$ simulation
are reweighted using the \mela package to model \offshell \Hboson production distributions, 
as an alternative approach to the direct \offshell simulation discussed above. The two approaches are
complementary and allow us to apply mutual cross-checks. Both approaches use the
same LO matrix elements available in the \mela package based on the \jhugen framework. 
Reweighting of simulated high-mass $\POWHEG + \jhugen$ events allows the modeling to start from NLO calculations 
of the SM signal hard-scattering production integrated with the parton shower simulation, 
compared to integrating LO production of direct \offshell samples.
This gives better modeling of the associated jet activity in the events. 
However, reweighting based on the LO matrix elements would lead to more complexity and internal 
inconsistencies in the QCD order, when applied to obtain both the background and interference predictions
from the signal NLO simulation. 

In the gluon fusion process, the factorization and renormalization scales are allowed to run by equating them to $\mell/2$.
To include higher-order QCD corrections, signal cross section calculations
are performed at LO, NLO, and next-to-NLO (NNLO), using the \MCFM and \hnnlo~2 programs~\cite{Catani:2007vq,Grazzini:2008tf,Grazzini:2013mca} 
with a narrow-width approximation, for a wide range of masses covering both on- and off-shell ranges. 
The ratios between the NNLO and LO cross section values, called the NNLO-to-LO $K$ factors, 
are used to reweight~\cite{deFlorian:2016spz} the \mell distributions from the \MCFM and \jhugen simulations at LO in QCD.
A uniform $K$ factor of 1.10 is applied across the entire \mell range to normalize the \Hboson production cross section via gluon fusion
to the predictions for $\mell\approx125\GeV$ at next-to-NNLO (N$^3\text{LO}$) in QCD~\cite{deFlorian:2016spz}.
The \mell distributions obtained from the $\POWHEG + \jhugen$ simulation of the \ggH process are reweighted 
with the NNLO-to-NLO $K$ factors. 

While the NNLO-to-LO $K$ factor calculation is directly applicable to the signal cross section,
it is only approximate for the $\Pg\Pg \to4\Pell$ background and its interference with the signal. 
An approximate NLO calculation~\cite{Caola:2015psa,Melnikov:2015laa,Campbell:2016ivq,Caola:2016trd} 
is available for the background and the interference. 
The resulting NLO-to-LO $K$ factors for the background and interference are consistent with that for the
signal within approximately 10\% 
in the mass range of $\mell > 220\GeV$ relevant for this analysis.
We therefore multiply the background and interference contributions by the same NNLO-to-LO $K$ factor 
and uniform N$^3\text{LO}$ correction, both calculated for the signal and including associated uncertainties. 
We introduce an additional 10\% uncertainty associated with this factor for the background 
and the square root of this variation ($\approx$5\%) for the interference.

The $\qqbar\to4\Pell$ background simulation is performed at NLO in QCD and LO in the EW theory with \POWHEG. 
The fully differential cross section for this process has been computed at NNLO in QCD~\cite{Grazzini:2015hta}, 
and the NNLO-to-NLO $K$~factor as a function of $m_{4\Pell}$ has been applied to the \POWHEG sample. 
This $K$ factor is 1.1 at $m_{4\Pell} = 125 \GeV$ and it varies from 1.0 to 1.2 in the $m_{4\Pell} < 500\GeV$ range. 
The uncertainty due to missing EW corrections in the region $m_{4\Pell} < 2m_{\PZ}$ is expected to be small 
compared to the uncertainties in the QCD calculation.
An EW NLO-to-LO $K$~factor~\cite{Bierweiler:2013dja} is applied to two \onshell \PZ bosons.
This $K$~factor depends on $m_{4\Pell}$, decreasing from unity (close to 125\GeV) to ${\approx}$0.9 (in the $m_{4\Pell} > 500\GeV$ range). 
The uncertainty in the latter is the dominant uncertainty for the off-shell width measurement 
and is applied as a function of $m_{4\Pell}$.
In the \onshell region, the EW background from the $\PV\PV\PV$, $\ttbar\PV\PV$, and $\ttbar\PV$ processes
is generated with~\MGvATNLO~\cite{Alwall:2014hca}.

All signal and background event generators are interfaced with the \PYTHIA~8.320 program~\cite{Sjostrand:2014zea} to simulate 
multiparton interactions, parton showering, and hadronization.
The CUETP8M1 and CP5 tunes~\cite{Khachatryan:2015pea,Sirunyan:2019dfx} 
are applied in simulating the 2016 and 2017--2018 data-taking periods, respectively.
The NNPDF~3.1 parton distribution functions~\cite{Ball:2011uy} are used for all simulated samples.
Simulated events include the contribution from additional $\Pp\Pp$ interactions within the same or adjacent
bunch crossings (pileup) and are weighted to reproduce the pileup distribution observed in data.
The generated events are further processed through a detailed simulation of the CMS detector based on  \GEANTfour~\cite{Agostinelli2003250}.

\section{Event reconstruction and selection}
\label{sec:EventSelection}
The event reconstruction is based on the particle-flow (PF) algorithm~\cite{Sirunyan:2017ulk}. This algorithm aims to reconstruct and identify each individual particle in an event, with an optimized combination of information from the various elements of the CMS detector. The energy of photons is obtained from the ECAL measurement. The energy of electrons is determined from a combination of the electron momentum at the primary vertex as determined by the tracker, the energy of the corresponding ECAL cluster, and the energy sum of all bremsstrahlung photons spatially compatible with originating from the electron track. The energy of muons is obtained from the curvature of the corresponding track. The energy of charged hadrons is determined from a combination of their momentum measured in the tracker and the matching ECAL and HCAL energy deposits, corrected for the response function of the calorimeters to hadronic showers. Finally, the energy of neutral hadrons is obtained from the corresponding corrected ECAL and HCAL energies.

Muons with $\pt > 5\GeV$ are reconstructed within the geometrical acceptance, corresponding to the region $\abs{\eta}< 2.4$, by combining information from the silicon tracker and the muon systems \cite{MuReco}. The muons are selected among the reconstructed muon track candidates by applying quality requirements on the track in both the muon and silicon tracker systems, and demanding small energy deposits in the calorimeters. A relative muon isolation variable is defined as
\ifthenelse{\boolean{cms@external}}{
  \begin{multline} \label{eqn:pfiso}
    {\mathcal I} \equiv \Big( \sum \pt^\text{charged} + \max\big[ 0, \sum \pt^\text{neutral} +\\
      \sum \pt^{\PGg} - \pt^{\mathrm{PU}} \big] \Big) / \pt^{\PGm} .
  \end{multline}
}
{
\begin{equation} \label{eqn:pfiso}
  {\mathcal I}^{\PGm} \equiv \Big( \sum \pt^\text{charged} + \max\big[ 0, \sum \pt^\text{neutral} +
    \sum \pt^{\PGg} - \pt^{\mathrm{PU}} \big] \Big) / \pt,
\end{equation}
}
where $\pt$ is the muon transverse momentum, and $\pt^{\text{charged}}$, $\pt^\text{neutral}$, and $\pt^\PGg$ are the transverse momentum of charged hadrons, neutral hadrons, and photons, respectively, within a cone radius of $\Delta R=0.3$ around the muon direction at the primary vertex.\\
Here, $\Delta R(i,j) \equiv \sqrt{\smash[b]{(\eta^i-\eta^j)^{2} + (\phi^i-\phi^j)^{2}}}$, where $\phi$ is the azimuthal angle, and in this case $\textit{i}$ refers to the hadron or photon, and $\textit{j}$ to the muon.
Since the isolation variable is particularly sensitive to energy deposits from pileup interactions, a contribution $\pt^{\mathrm{PU}}$ from pileup is subtracted from the isolation parameter, as shown in Eq.~(\ref{eqn:pfiso}). 
It is defined as 0.5 times the \pt sum of all charged hadrons $i$ not originating from the primary vertex $\pt^{\mathrm{PU}} = 0.5\sum_{i}\pt^{i,\mathrm{PU}}$, where the 0.5 factor accounts for the different fraction of charged and neutral hadrons~\cite{PUmitigationCMS}. A requirement of ${\mathcal{I}^{\PGm}} < 0.35$ is placed on each muon in the event.

Photons from final-state radiation are reconstructed using the PF algorithm~\cite{Sirunyan:2017ulk}. Isolated photons with $\pt >2\GeV$, $\abs{\eta} <2.4$, and $\mathcal{I}^{\PGg} < 1.8$, are associated with the closest lepton (either muon or electron) in the event. Photons that do not satisfy the requirements $\Delta R(\PGg,\Pell)/(\pt^{\PGg})^2<0.012\GeV^{-2}$ and $\Delta R(\PGg,\Pell) < 0.5$ are discarded.
If more than one photon candidate fulfils the above conditions, the one with the lowest value of $\Delta R(\PGg,\Pell)/(\pt^{\PGg})^2$ with respect to the given lepton is retained.
Photons passing the above criteria are excluded from the computation of the relative isolation parameter. 

Electrons with $\pt>7\GeV$ are reconstructed within the geometrical acceptance, corresponding to the pseudorapidity region $\abs{\eta}<2.5$~\cite{eleReco}. They are identified using a multivariate discriminant, which includes observables sensitive to the emission of bremsstrahlung along the electron trajectory, the geometrical and momentum-energy matching between the electron trajectory and the associated cluster in the ECAL, the shape of the electromagnetic shower in the ECAL, and variables that discriminate against electrons originating from photon conversions. The isolation parameter sums for electrons, defined similarly as for muons, are included in the multivariate discriminant. This information is used to discriminate between prompt leptons from \PZ boson decays and those arising from EW decays of hadrons within jets. The package \textsc{XGBoost}~\cite{xgboost} is used to train and optimize this multivariate discriminant. The training is performed with simulated events that are not used at any other stage of the analysis. Separate trainings are performed for the three different data-taking periods \cite{ElectronBDT}. Electrons from photon conversions and muons from in-flight decays of hadrons are rejected if the ratio of their impact parameter in three dimensions, computed with respect to the primary vertex position, to their uncertainty is greater than four.

The reconstruction and selection efficiencies for prompt leptons in both data and simulation have been estimated using a tag-and-probe technique~\cite{CMS:2011aa} based on samples of $\PZ$ boson events. The ratio of the efficiencies measured in data and simulation is used to rescale the yields of selected events in the simulated samples.
In addition, $\PZ$ boson events have been used to calibrate the momentum scale and resolution of electrons and muons in bins of different kinematic variables~\cite{ScaleSmear2, Roch2}.

In the selection of Higgs boson candidates,
four prompt and isolated leptons are required following the prescription above. 
One of the leptons must have \pt$>20$\GeV and at least one of the remaining leptons must satisfy $\pt>10\GeV$.
The \PZ boson candidates are constructed from $\Pep\Pem$ or $\PGmp\PGmm$ pairs whose  invariant mass is in the range 12--120\GeV. 
The dilepton pairs are then combined to form the Higgs boson candidate. 
In the following, $\PZ_1$ denotes the pair with the mass closest to the nominal $\PZ$ boson mass, 
while $\PZ_2$ refers to the remaining one. Four possible combinations are considered and treated separately: 
4\PGm, 4\Pe, $2\Pe2\PGm$, and $2\PGm2\Pe$, where the mixed-flavor final states are separated based on the decay of $\PZ_{1}$. 
The four possible combinations have different four-lepton mass resolutions (largely driven by whether the $\PZ_{1}$ is formed from 2$\PGm$ or 2$\Pe$) 
and different amounts of reducible background (largely driven by whether the $\PZ_{2}$ decays to 2\PGm or 2\Pe).
None of the above differences between the flavor channels affect the \offshell \Hboson analysis 
and therefore all flavor channels are combined in that measurement. 
Signal candidates must satisfy $m_{4\Pell} > 70$\GeV. 
If more than one \Hboson candidate can be formed in the event, the one with the highest value of the 
kinematic discriminant $\mathcal{D}_\text{bkg}^\text{kin}$, defined in Section~\ref{sec:Discriminants},
is retained, unless these candidates consist of the same four leptons. 
In this case, the candidate with the $\PZ_1$ invariant mass closest to the nominal \PZ boson mass is retained.

In the \offshell analysis, events are further categorized based on the jets associated with the \Hboson candidate. 
The jets are clustered using the anti-\kt jet finding algorithm~\cite{Cacciari:2008gp,Cacciari:2011ma} with a distance 
parameter of 0.4. The jet momentum is determined as the vector sum of all particle momenta in the jet. 
Jets must satisfy $\pt>30\GeV$ and $\abs{\eta}<4.7$ and must be separated
from all selected lepton candidates and any selected final-state radiation photons by demanding 
$\Delta R(\Pell/\cPgg,{\text{jet}})>0.4$.
Jets originating from the hadronization of \PQb quarks are identified using the DeepCSV algorithm~\cite{Sirunyan:2017ezt}, which combines information 
on impact parameter significance, the secondary vertex, and jet kinematic variables. The use of this identification is described in Section~\ref{sec:Discriminants}.

\section{Lepton momentum measurement and four-lepton mass resolution}
\label{sec:LepImprov}
Once an event is selected in the \Hboson mass and \onshell width measurements, beam spot information is used to improve the $m_{4\Pell}$ resolution.
In this approach (denoted as BS in the following), the beam spot position is included as a common point in the reconstruction of the tracks from the Higgs boson decay. 
While muon kinematic parameters such as \pt are recalculated after implementing this constraint,
electron kinematic parameters are not modified. 
This is because the electron momentum is determined mostly by the energy measurement in the ECAL, rather than from the track curvature. Thus, applying the beam spot constraint does not change the momentum measurement.
In the final states involving muons, the BS constraint improves the $m_{4\Pell}$ resolution by about 3--8\%, mostly depending on the flavor of the leptons originating from $\PZ_1$.

Once the BS approach is applied, an important variable, the event-by-event four-lepton mass uncertainty ($\delta m_{4\Pell}$), is introduced.
Individual lepton momentum uncertainty (per-lepton uncertainty in the following) is obtained from the algorithms used to estimate lepton momenta. For muons, the full covariance matrix is obtained from the muon track fit, and the uncertainty in the muon direction is assumed to be negligible. For the electrons, the momentum uncertainty is estimated from the combination of the ECAL and silicon tracker measurements, neglecting the uncertainty in the track direction. The per-lepton momentum uncertainty is then propagated to the $m_{4\Pell}$ to predict $\delta m_{4\Pell}$.
Each $\delta m_{4\Pell}^i$ corresponding to individual lepton momentum variation is calculated separately. The total per-event uncertainty $\delta m_{4\Pell}$ is obtained as the quadrature sum of the four individual $\delta m_{4\Pell}^i$.  

The accuracy of the $\delta m_{4\Pell}$ determination is improved by applying correcting factors to the momentum uncertainty. These corrections are derived for muons in
several $\abs{\eta}$ bins, and for electrons in bins of $\delta$\pt/\pt vs. $\abs{\eta}$. 
The $2\Pell$ invariant mass distribution is used to extract these corrections. The mass distribution is fit in two steps, to ensure a more reliable result, with a convolution of a Breit--Wigner function (that describes the intrinsic shape of the \PZ boson resonance) with a double-sided Crystal Ball function~\cite{Skwarnicki:1986xj} (to describe the detector effects), plus an exponential function (used to describe the background). 
The first step is used to estimate all the parameters in the fit function but not its resolution. 
In the second step, the only floating parameter of the fit is the resolution, 
after fixing all the others to the values extracted before. From this second fit, 
the resolution and the correction to the momentum uncertainty are determined.
After the corrections are extracted, a closure test between the predicted and measured invariant mass resolutions is performed.

To further improve the $m_{4\Pell}$ resolution, a kinematic fit is also performed
using a mass constraint on the intermediate \onshell \PZ resonance, with an approach similar to the one
described in Ref.~\cite{Sirunyan:2017exp}. 
For a 125 \GeV \Hboson, the selected $Z_{1}$ is mostly \onshell,
while the $m_{\PZ_{2}}$ invariant mass distribution is broad and the width is much larger than the detector resolution.
When considering the \Hboson mass measurement, the expected gain in resolution comes from refitting $Z_{1}$.
Thus, it is possible to reevaluate the \pt of the two leptons forming
$\PZ_{1}$, using the constraint that the reconstructed invariant mass of the \PZ boson candidates must follow
the true shape of the \PZ boson. 
In every event, the lepton momenta are adjusted using the likelihood function
  \ifthenelse{\boolean{cms@external}}{
    \begin{multline*}
      \mathcal{L}(\pt^{1},\pt^{2}\vert\pt^\text{reco1},\sigma^{1},\pt^\text{reco2},\sigma^{2}) = \text{Gauss}(\pt^\text{reco1}\vert \pt^{1},\sigma^{1}) \\
      \text{Gauss}(\pt^\text{reco2}\vert\pt^{2},\sigma^{2}) \mathcal{L}(m_{12}\vert m_{\PZ},m_{\PH}),
    \end{multline*}
  }{
    \begin{equation*}
      \mathcal{L}(\pt^{1},\pt^{2}\vert \pt^\text{reco1},\sigma^{1},\pt^\text{reco2},\sigma^{2}) =
      \text{Gauss}(\pt^\text{reco1}\vert \pt^{1},\sigma^{1}) \text{Gauss}(\pt^\text{reco2}\vert \pt^{2},\sigma^{2})
      \mathcal{L}(m_{12}\vert m_{\PZ},m_{\PH}),
    \end{equation*}
  }
where $\pt^\text{reco1}$ and $\pt^\text{reco2}$ are the reconstructed transverse momenta of the two leptons forming the $Z_{1}$,
$\sigma^1$ and $\sigma^2$ are the per-lepton resolutions (uncertainties in the \pt measurements, corrected using the method described above), $\pt^{1}$ and $\pt^{2}$ are the observables under optimization, and $m_{12}$ is the invariant mass calculated from $\pt^{1}$ and $\pt^{2}$. 
Finally, $\mathcal{L}(m_{12}\vert m_{\PZ},m_{\PH})$ is the likelihood function given the true shapes of $m_{\PZ_{1}}$ and of the 125\GeV \Hboson.
For each event, the likelihood is maximized and the \pt values of the refitted leptons are updated.

\begin{figure*}[!htb]
    \centering
    \includegraphics[width=0.42\textwidth]{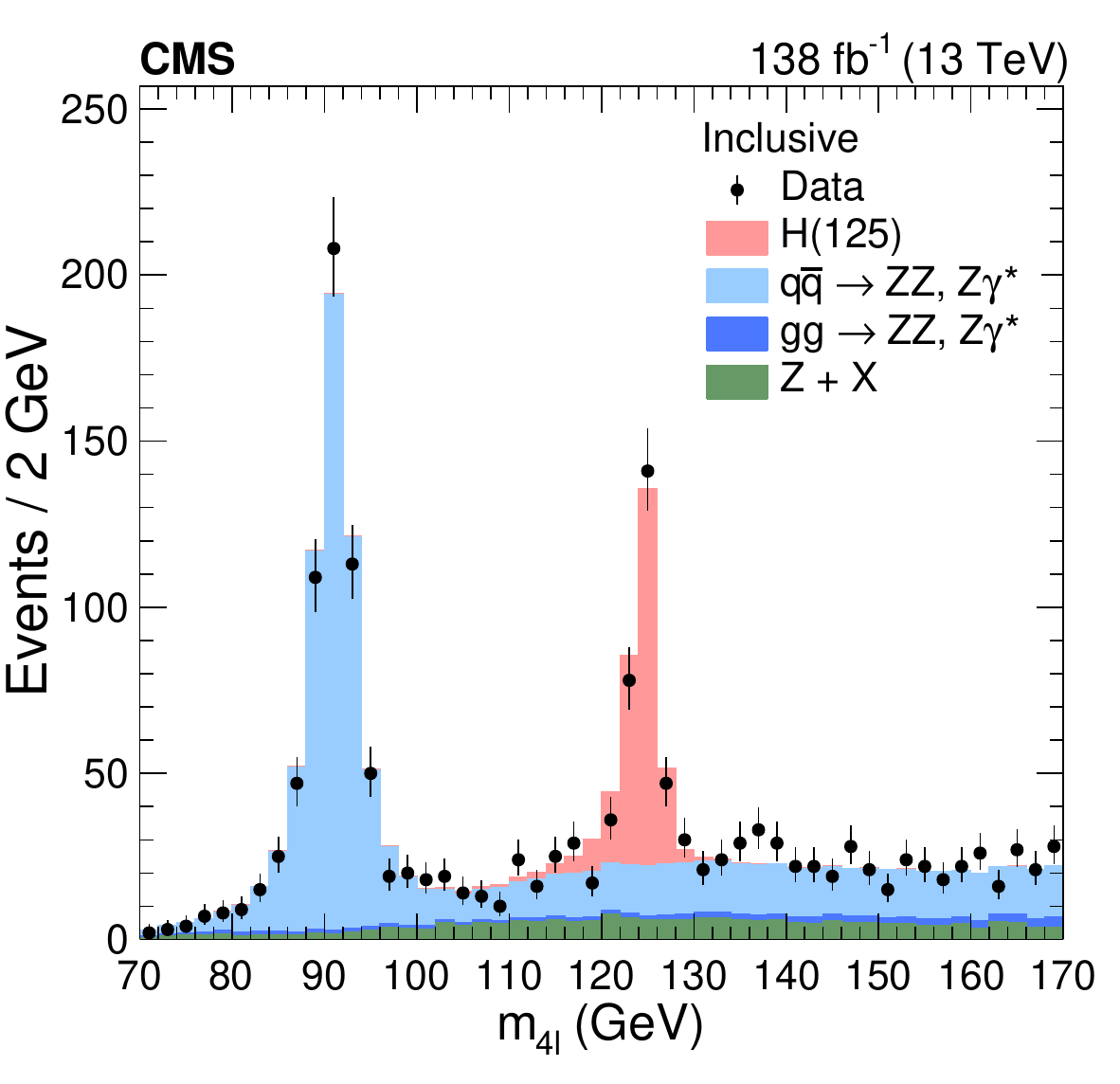}\\
    \includegraphics[width=0.42\textwidth]{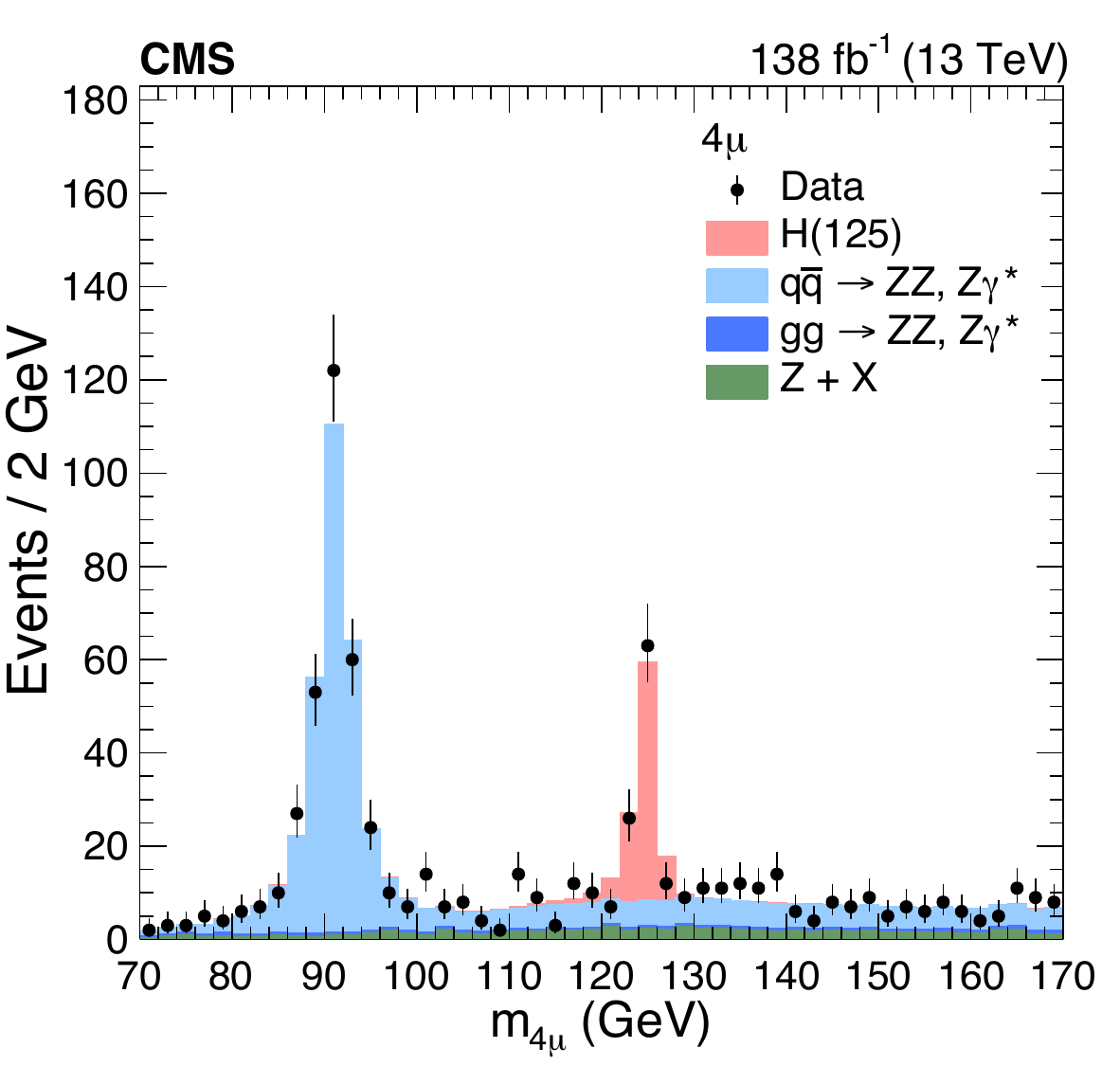}
    \includegraphics[width=0.42\textwidth]{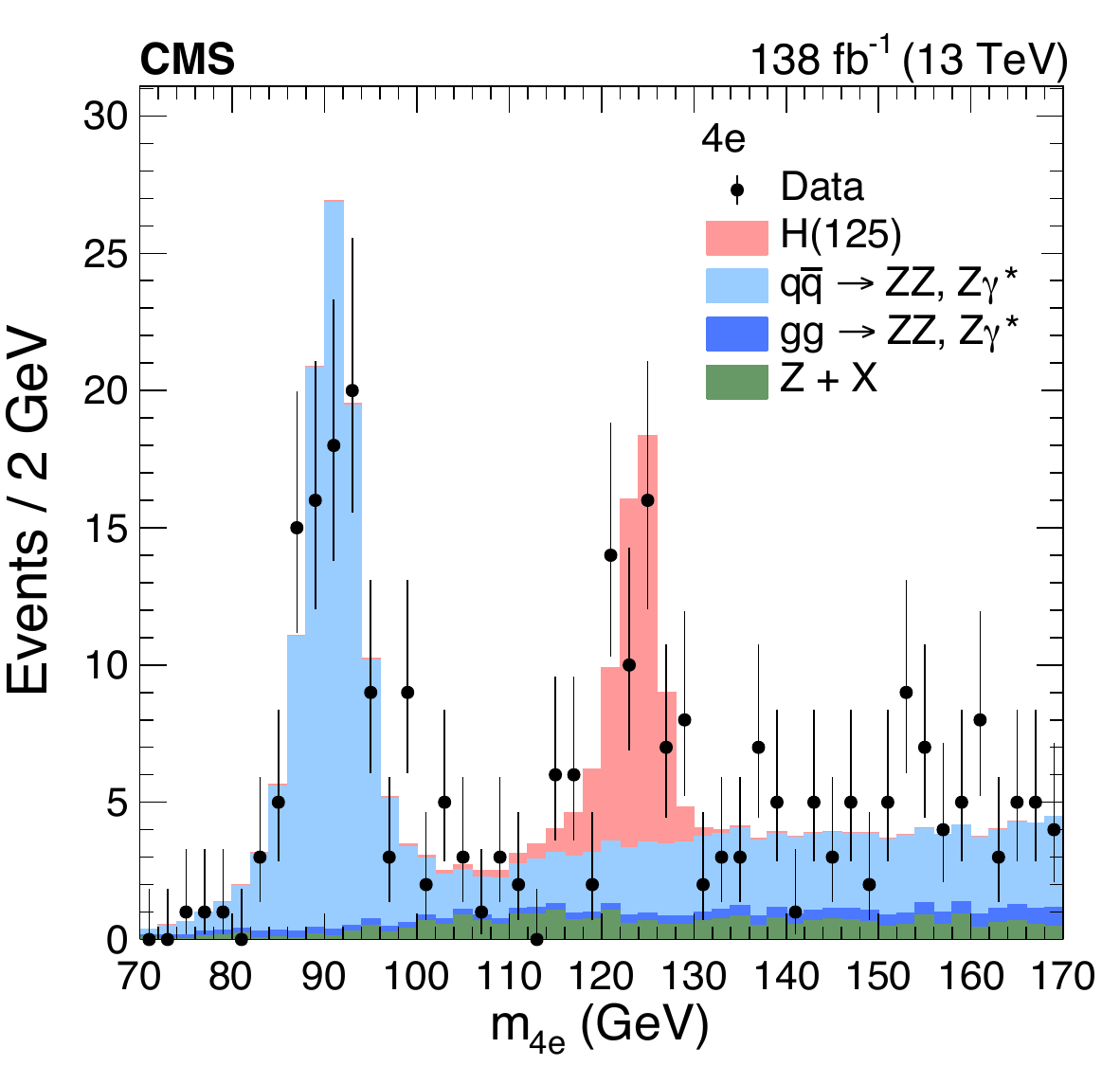}\\
    \includegraphics[width=0.42\textwidth]{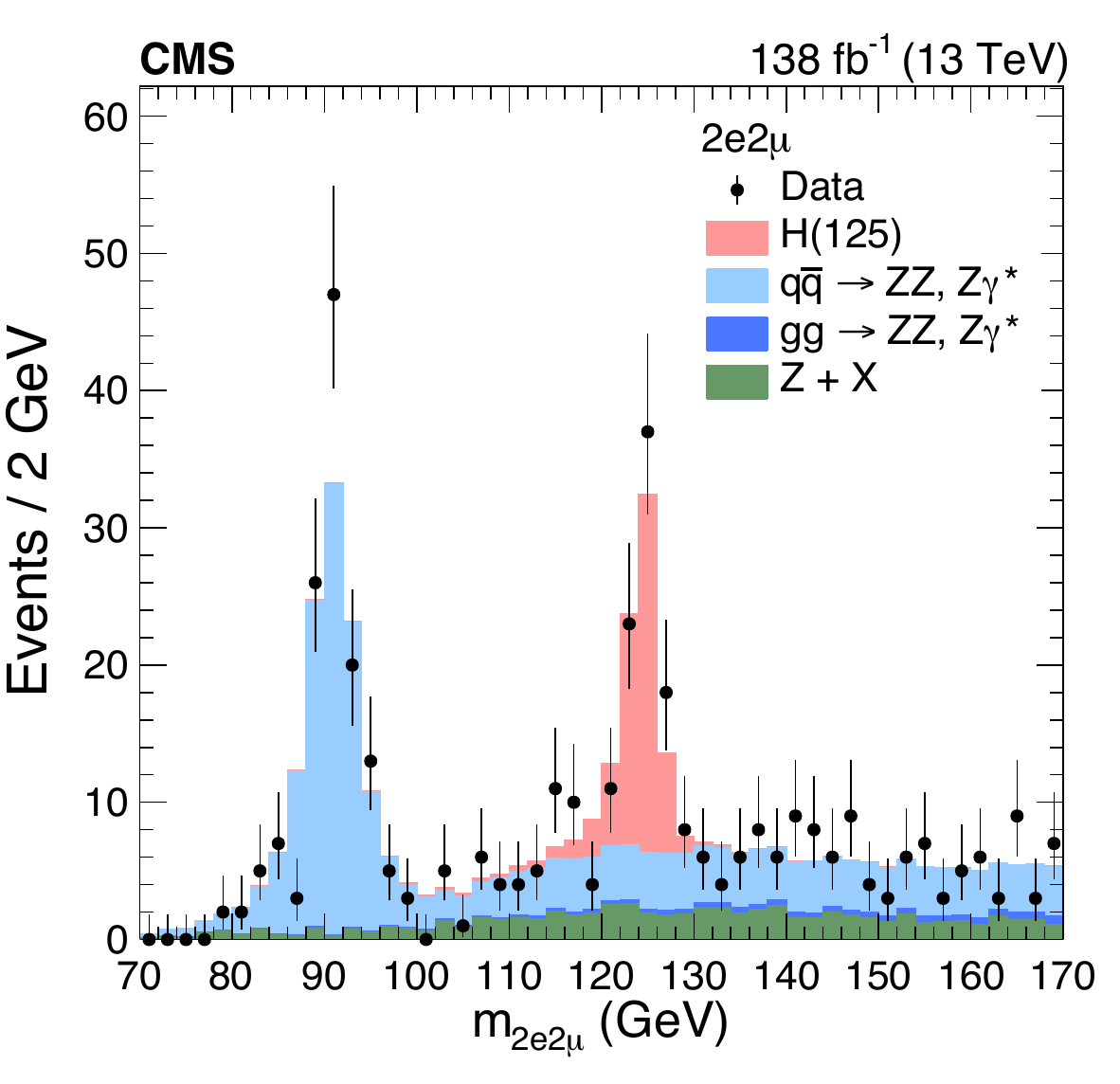}
    \includegraphics[width=0.42\textwidth]{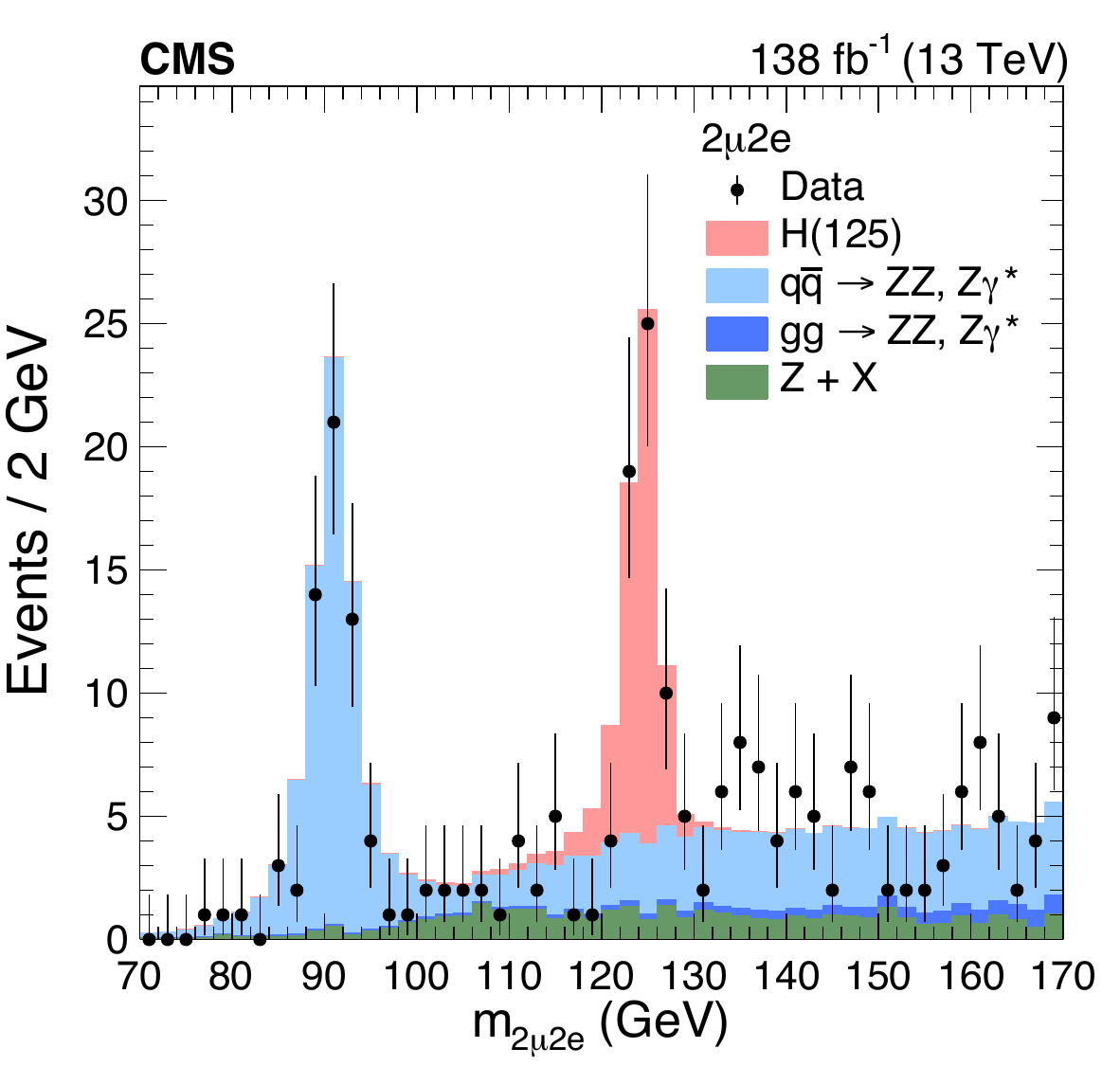}\\
    \caption{The observed (points) and predicted (stacked histograms) $m_{4\Pell}$ distributions in the inclusive (upper), 4$\PGm$ (middle left), 4$\Pe$ (middle right), 2$\Pe$2$\PGm$ (lower left) and 2$\PGm2\Pe$ (lower right) final states, defined such that the first lepton pair is taken to be the one with the mass closest to the nominal $\PZ$ boson mass. The predictions for the \Hboson signal and the three main backgrounds are given by the different colors.  The vertical bars on the points show the statistical uncertainties in the data.}
    \label{Mass_distributions}
  \end{figure*}

\begin{figure}[!htb]
  \centering
  \includegraphics[width=0.45\textwidth]{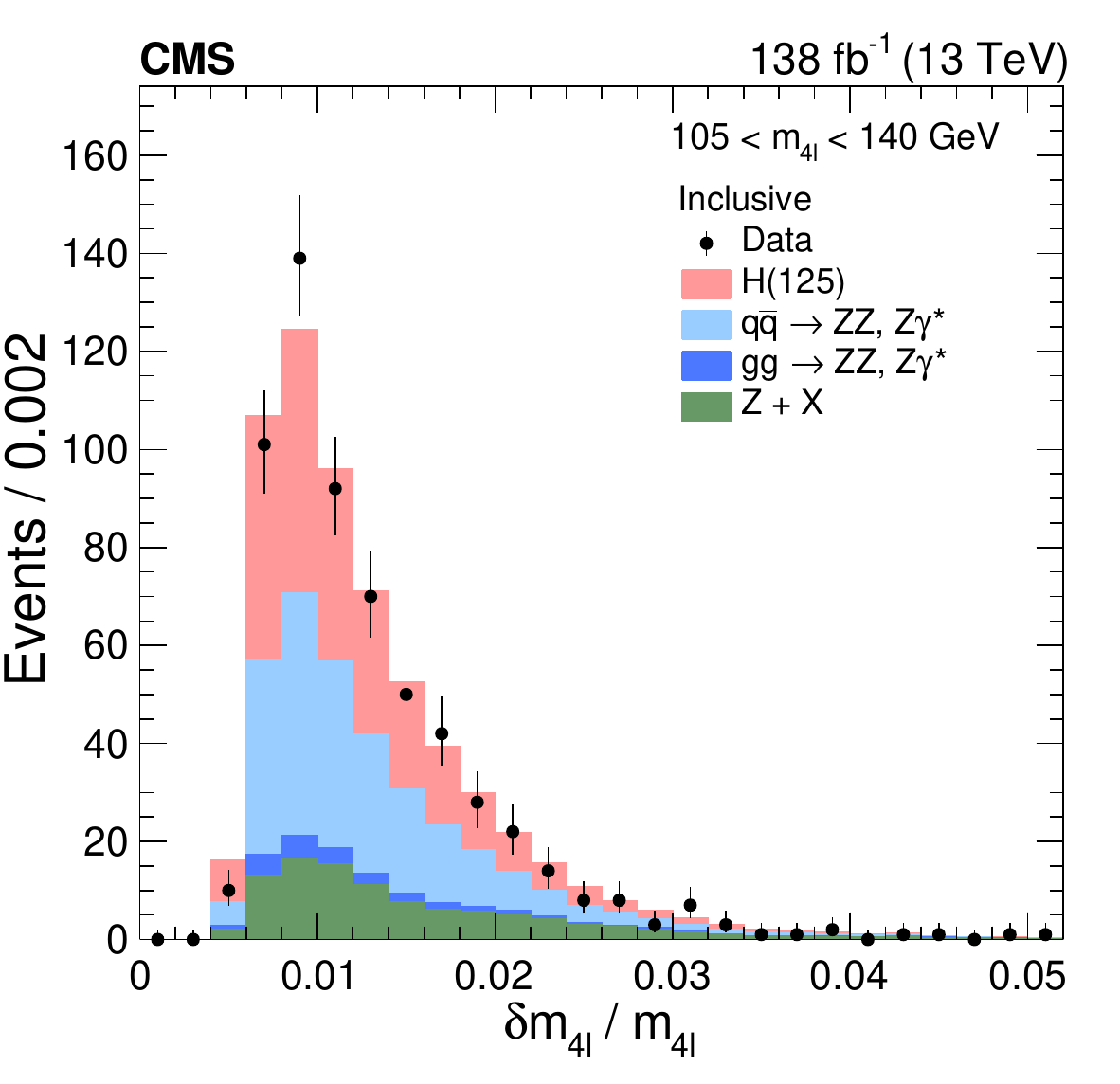}
  \caption{Distributions of the observed (points) and predicted (stacked histograms) relative per-event mass
    uncertainty of the four-lepton system, in the inclusive final state. The predictions for the \Hboson signal and the three main backgrounds are given by the different colors.  The vertical bars on the points show the statistical uncertainties in the data.}
  \label{MassErr_distribution}
\end{figure}

\begin{figure}[!htb]
  \centering
  \includegraphics[width=0.45\textwidth]{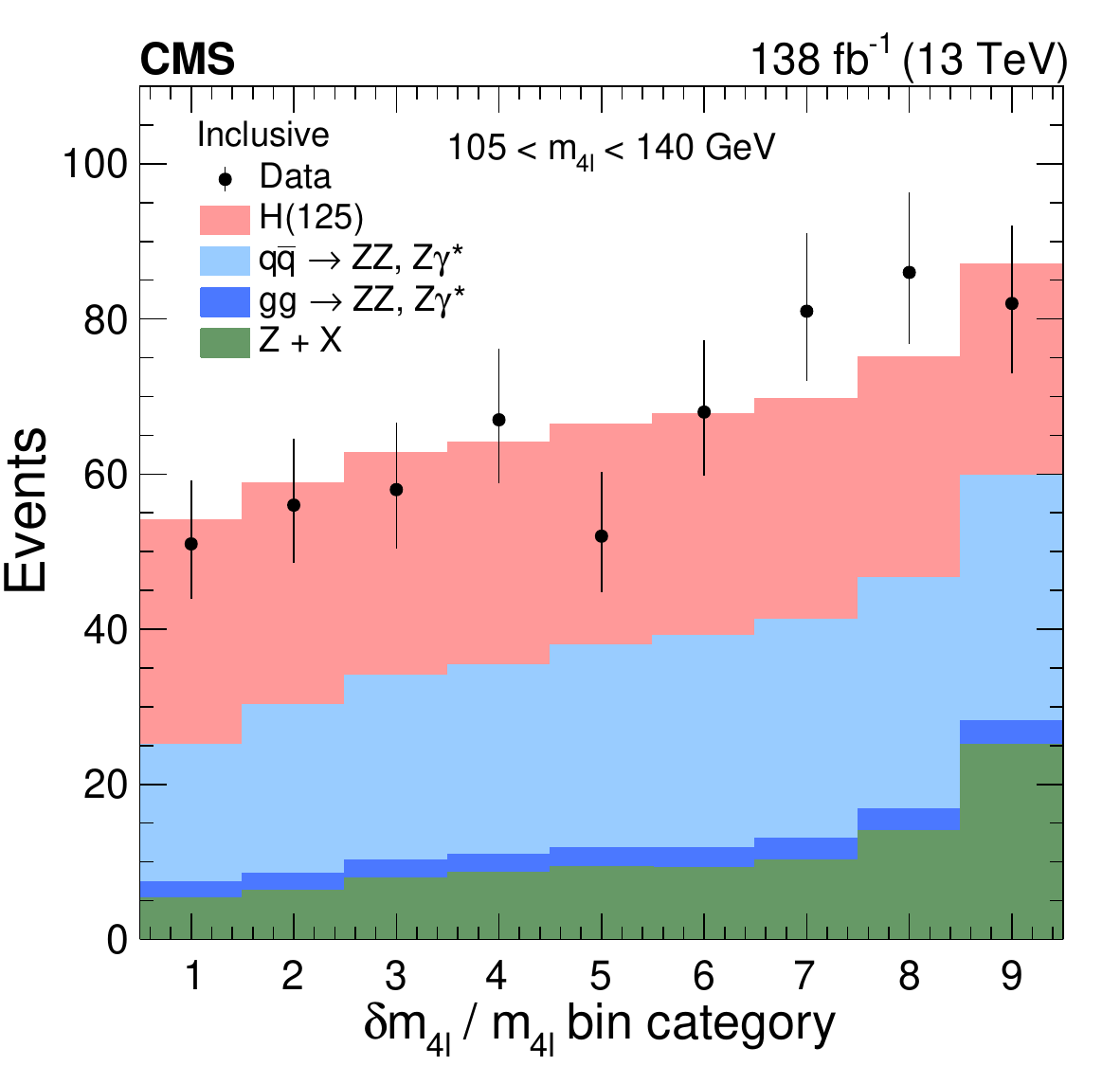}
  \caption{Observed (points) and predicted (stacked histograms) yields of four-lepton events in each of the 9 \RelMassErr bin categories for the inclusive final state.
    The bins are shown in the order of increasing \RelMassErr. The predictions for the \Hboson signal and the three main backgrounds are given by the different colors.  The vertical bars on the points show the statistical uncertainties in the data.}
  \label{DataMC_category}
\end{figure}

For the \Hboson \offshell production, the $m_{4\Pell}$ distribution is much broader than the $m_{4\Pell}$ resolution, 
and, therefore, improvements in the treatment of the lepton momentum uncertainties are not applied.
Figure~\ref{Mass_distributions} shows the $m_{4\Pell}$ distribution inclusively for all the four-lepton states (upper plot) and separately for the four individual final states. 
Expectations for the signal and background contributions are shown as stacked histograms. The yields from the signal and $\PZ\PZ$ backgrounds are estimated from simulation assuming SM cross sections, while the $\PZ+\PX$ background yield is estimated from data. More details are given in Sections~\ref{Bkg2} and~\ref{SignalModelling}.
Figure~\ref{MassErr_distribution} displays the distribution of the relative per-event mass uncertainty of the four-lepton system (\RelMassErr) in the inclusive final state with $105< m_{4\Pell} <140 \GeV$. This variable is used to categorize events in the \onshell region according to their invariant mass resolution. The categorization is performed by defining nine mutually exclusive bins with an equal number of signal events in each. 
Since each final state has a different resolution that also varies across data-taking periods, the bin boundaries are determined independently per final state and per data-taking year.
Figure~\ref{DataMC_category} plots the observed and predicted yields of four-lepton events as a function of \RelMassErr bin for the $105< m_{4\Pell} <140$\GeV invariant mass range.
This approach improves the precision of the \Hboson mass measurement by about 10\%.

\section{Kinematic discriminants and off-shell event categorization}
\label{sec:Discriminants}
The reconstructed four-lepton events and their associated jets, where relevant, can be described 
with several observables using the kinematic features of the \Hboson decay and associated particles.
It is a challenging task to perform an optimal analysis in a large multidimensional phase space.
There are up to 13 observables in the set $\boldsymbol{\Omega}$ that describe the \Hboson kinematic distributions in the $2\to 6$ process
of collision of two protons and the production of four leptons and two associated jets.
The \mela approach is designed to reduce the number of observables to the minimum while retaining all essential information. 
In this method, optimal discriminators are defined through the utilization of matrix element likelihood 
calculations~\cite{Gao:2010qx,Bolognesi:2012mm,Anderson:2013afp,Gritsan:2016hjl,Gritsan:2020pib}.
Two types of discriminants are defined for each of the \Hboson production and decay processes as
\begin{equation}
\mathcal{D}_\text{alt}\left(\boldsymbol{\Omega}\right) = \frac{\mathcal{P}_\text{sig}\left(\boldsymbol{\Omega}\right) }
        {\mathcal{P}_\text{sig}\left(\boldsymbol{\Omega}\right) +\mathcal{P}_\text{alt}\left(\boldsymbol{\Omega}\right) }
\label{eq:melaD}
\end{equation}
and
\begin{equation}
\mathcal{D}_\text{int}\left(\boldsymbol{\Omega}\right) =
\frac{\mathcal{P}_\text{int}\left(\boldsymbol{\Omega}\right) }
{2 \ \sqrt{{\mathcal{P}_\text{sig}\left(\boldsymbol{\Omega}\right) \ \mathcal{P}_\text{alt}\left(\boldsymbol{\Omega}\right) }}},
\label{eq:melaDint}
\end{equation}
where the probability of a certain process $\mathcal{P}$ is calculated using the full set of kinematic observables 
$\boldsymbol{\Omega}$ for the processes denoted as ``sig'' for a signal model and ``alt'' for an alternative model,
which can be for either an alternative \Hboson production mechanism, or a background, depending on the individual case.
The ``int'' label represents the interference between the two model contributions.
The probabilities $\mathcal{P}$ are calculated from the matrix elements obtained from the \mela approach.

\begin{figure}[!htbp]
\centering
\includegraphics[width=0.45\textwidth]{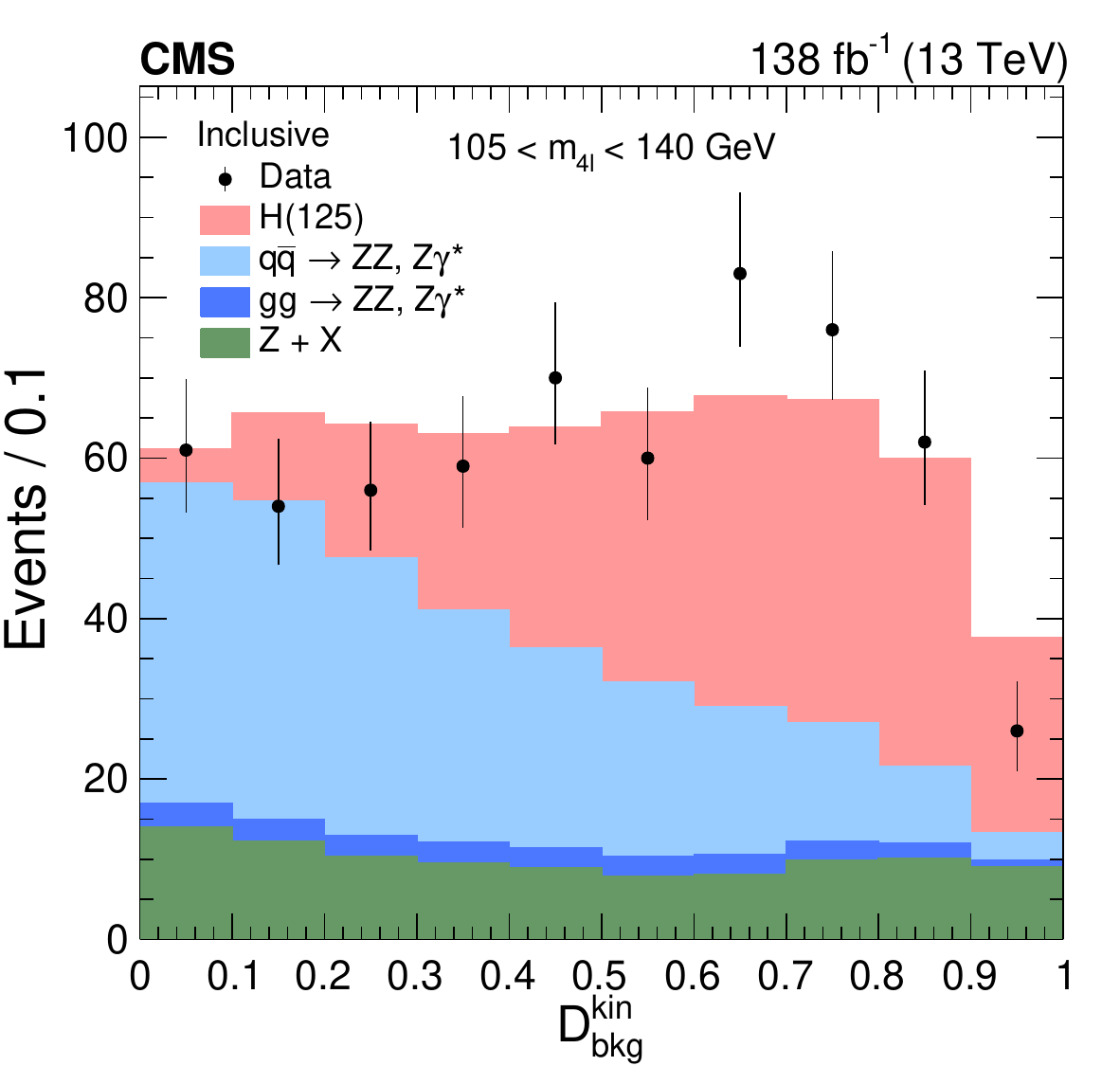}
\caption{Distributions of the observed (points) and predicted (stacked histograms) \Dbkgkin of the four-lepton system, in the inclusive final state. The predictions for the \Hboson signal and the three main backgrounds are given by the different colors.  The vertical bars on the points show the statistical uncertainties in the data.}
\label{Dbkgkin_distribution}
\end{figure}

The \Dbkgkin discriminant is used in the selection of events, as discussed in Section~\ref{sec:EventSelection},
and as one of the observables in the maximum likelihood fit for the signal extraction, in both the on- and off-shell analyses. 
It is determined following Eq.~(\ref{eq:melaD}),
where $\mathcal{P}_\text{bkg}$ is calculated for the dominant $\qqbar\to4\Pell$ background process
and $\mathcal{P}_\text{sig}$ is found for the $\PH\to4\Pell$ decay using the full kinematic information of the four leptons.
Figure~\ref{Dbkgkin_distribution} shows the inclusive \Dbkgkin distribution of  the four-lepton system. 
In the \offshell analysis, the \Dbkgkin discriminant is used for events without associated jets. 

For the two categories of \offshell events with two associated jets described below, \VBF- and \VH-tagged events, $\mathcal{P}_\text{bkg}$ and $\mathcal{P}_\text{sig}$ include the 
kinematic information on the four leptons and the two jets.
The $\mathcal{P}_\text{bkg}$ probability density corresponds to the EW and QCD background processes with four leptons and two jets,
while $\mathcal{P}_\text{sig}$ is determined for the EW signal processes \VBF and \VH. Including jet kinematic
information in the \Dbkg calculation improves the separation of the signal from both the background 
and \ggH production, when compared to \Dbkgkin.

In the analysis of the \offshell region, 
a dedicated study of a particular kinematic topology is done, treating gluon fusion and EW production separately, because they evolve differently with \mell, thus allowing us to probe their individual behaviors. 
Therefore, in the \offshell region, events are further split into three mutually 
exclusive categories based on the presence of other particles produced in association with the \Hboson candidate~\cite{Sirunyan:2021rug}.
This is not necessary in the on-shell analysis.

We use various kinematic discriminants and other selection requirements to perform the categorization of the \offshell events. 
The definition of these discriminants is given by Eq.~(\ref{eq:melaD}).
They are specifically designed to distinguish the targeted signal production mechanisms (\VBF, \WH, \ZH) 
within each category from gluon fusion production and are denoted as
$\mathcal{D}_\text{2jet}^{\VBF}$, $\mathcal{D}_\text{2jet}^{\PZ\PH}$, and $\mathcal{D}_\text{2jet}^{\PW\PH}$. 
They are labeled to indicate a specific dijet topology (``2jet'')  and production mechanism (\VBF, \WH, \ZH). 
Calculated using the \mela approach with matrix elements at LO in QCD, these discriminants utilize the complete 
kinematic information from the \Hboson decay and the associated jets. 
Additional details can be found in Refs.~\cite{Sirunyan:2017exp,Sirunyan:2017tqd,Sirunyan:2019twz,Sirunyan:2021rug}. 

The selected \offshell events are split into three categories: \VBF-tagged, \VH-tagged, and Untagged events,
as summarized in Table~\ref{tab:categoriesoffshell}. 
The discriminants $\mathcal{D}_\text{2jet}$ are constructed following Eq.~(\ref{eq:melaD}),
where $\mathcal{P}_\text{sig}$ corresponds to the signal probability for \VBF ($\WH$ or $\ZH$)
production in the \VBF-tagged (\VH-tagged) category, and $\mathcal{P}_\text{alt}$
to \Hboson production via \ggH in association with two jets.
When more than two jets pass the selection criteria, the two jets with the highest \pt are chosen 
for the matrix element calculations. Thereby, the $\mathcal{D}_\text{2jet}$ discriminants separate the 
targeted signal production mode of each category from \ggH production using 
only the kinematic variables of the \Hboson and the two associated jets.
Sequential selection criteria are applied to define the categories as follows:
\begin{itemize}
\item[--]  The {$\VBF$-2jet} category requires exactly four leptons. In addition, there must be 
either two or three jets of which at most one is identified as coming from the hadronization of a \PQb quark, which we term a \PQb-tagged jet~\cite{Sirunyan:2021rug}, 
or at least four jets and no \PQb-tagged jets. 
Finally, $\mathcal{D}_\text{2jet}^{\VBF} >0.5$ is required~\cite{Sirunyan:2021rug}. 
\item[--]  The {$\VH$-hadronic} category requires exactly four leptons. In addition, there must be
either two or three jets, or at least four jets and no \PQb-tagged jets. 
Finally, we demand $\max(\mathcal{D}_\text{2jet}^{\WH},~\mathcal{D}_\text{2jet}^{\ZH} )>0.5$~\cite{Sirunyan:2021rug}.
\item[--]  The {Untagged} category consists of the remaining events.
\end{itemize}
Table~\ref{tab:categoriesoffshell} provides a summary of the key categorization requirements,
with selections applied sequentially from left to right to define the three mutually exclusive categories.
All discriminants are calculated with the \jhugen signal and \MCFM background matrix elements. 
The \VH interference discriminant in the \VH-tagged category is defined as the simple average 
of the ones corresponding to the \ZH and \WH processes.
The use of decay kinematic information is denoted by the label ``dec''.

\begin{table*}[!hbtp]
\centering
\topcaption{
Summary of the three production categories in the \offshell \mell region and the observables used in the fits.
}
\label{tab:categoriesoffshell}
\begin{scotch}{lccc}
Category              & VBF-tagged & $\PV\PH$-tagged  & Untagged \\
\hline\\[-2ex]
Selection
 & $ \mathcal{D}_\text{2jet}^\text{VBF} >0.5$ 
 & $ \mathcal{D}_\text{2jet}^{\Z\PH}$ or $ \mathcal{D}_\text{2jet}^{\PW\PH}>0.5$
 & Rest of the events \\
  Observables
&  \mell, $\mathcal{D}^{\text{VBF}+\text{dec}}_\text{bkg}$, $\mathcal{D}_\text{bsi}^{\text{VBF}+\text{dec}}$
&  \mell, $\mathcal{D}^{\PV\PH+\text{dec}}_\text{bkg}$, $\mathcal{D}_\text{bsi}^{\VH+\text{dec}}$
&  \mell, $\mathcal{D}^\text{kin}_\text{bkg}$, $\mathcal{D}_\text{bsi}^{{\Pg\Pg},\text{dec}}$  \\
\end{scotch}
\end{table*}

\begin{figure*}[!htb]
\centering
\includegraphics[width=0.31\textwidth]{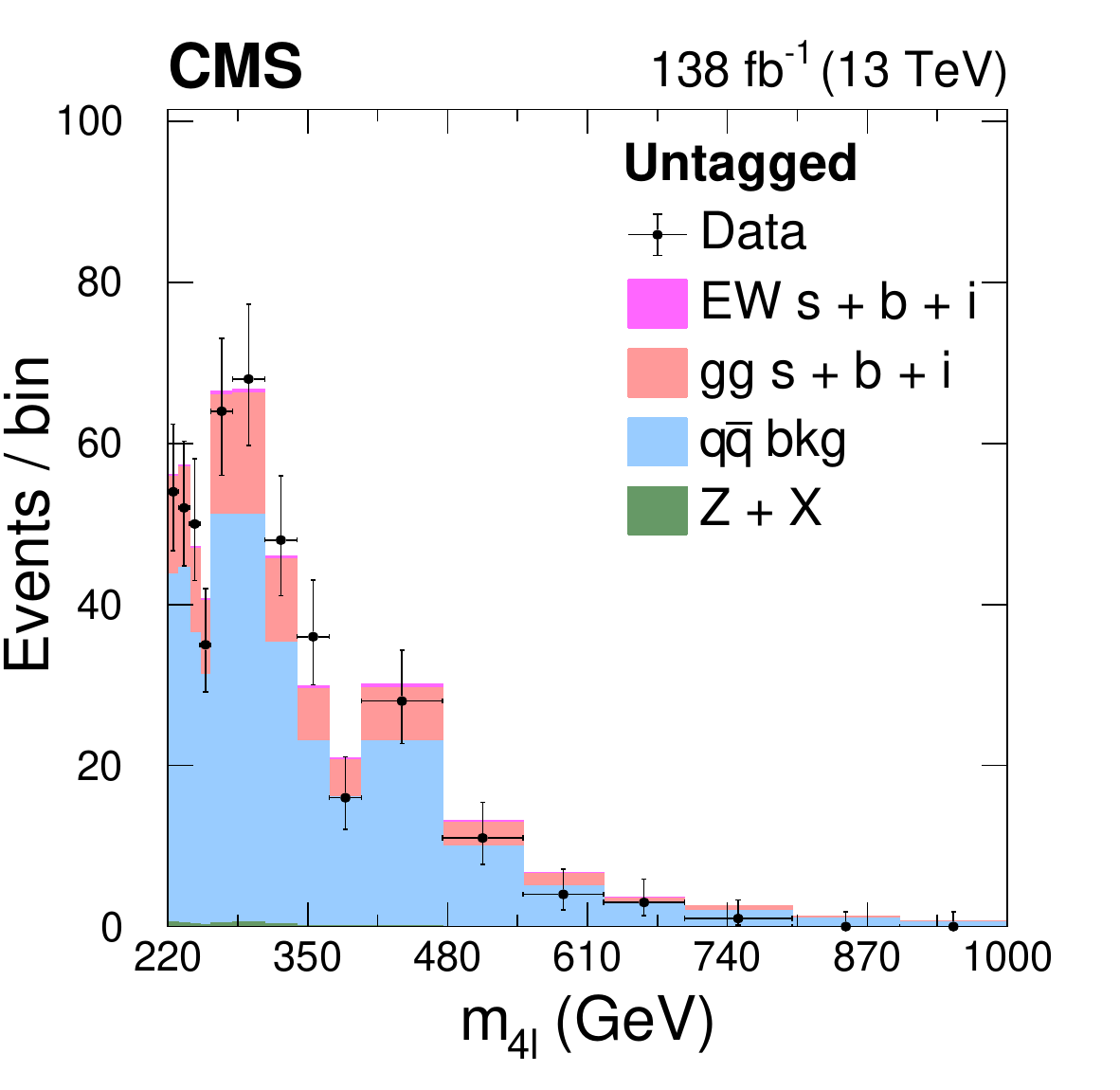}
\includegraphics[width=0.31\textwidth]{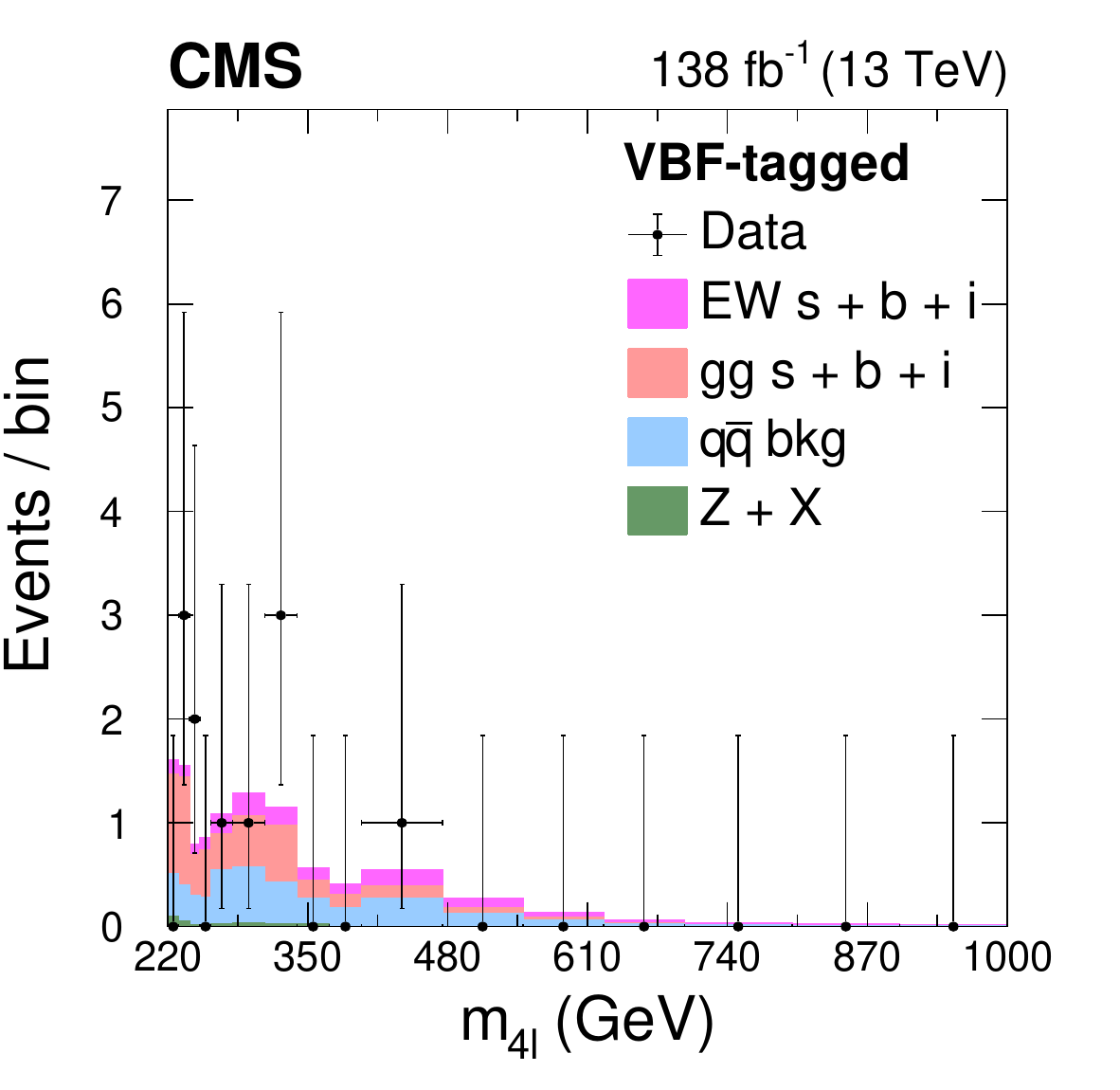}
\includegraphics[width=0.31\textwidth]{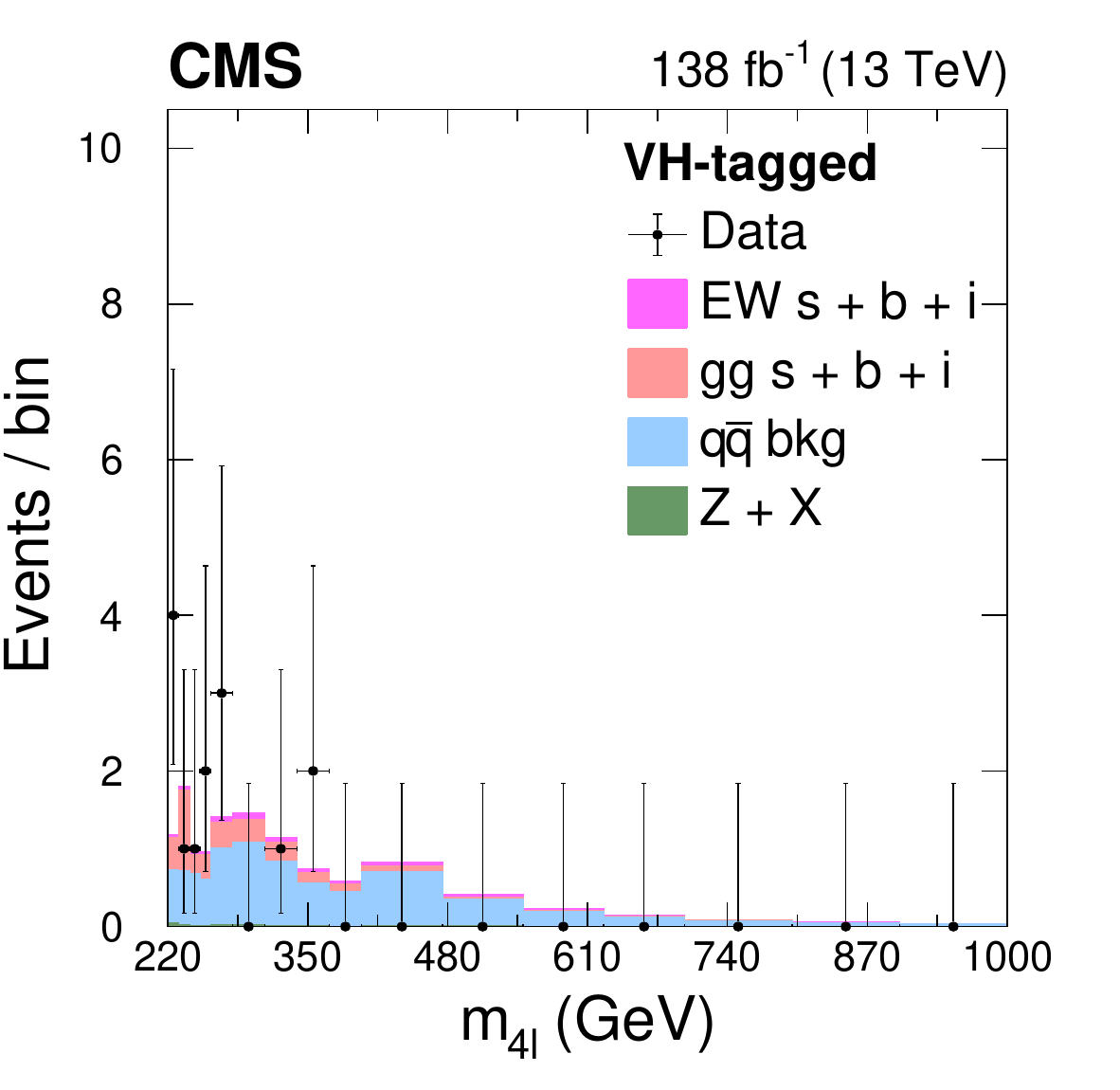}\\ 
\includegraphics[width=0.31\textwidth]{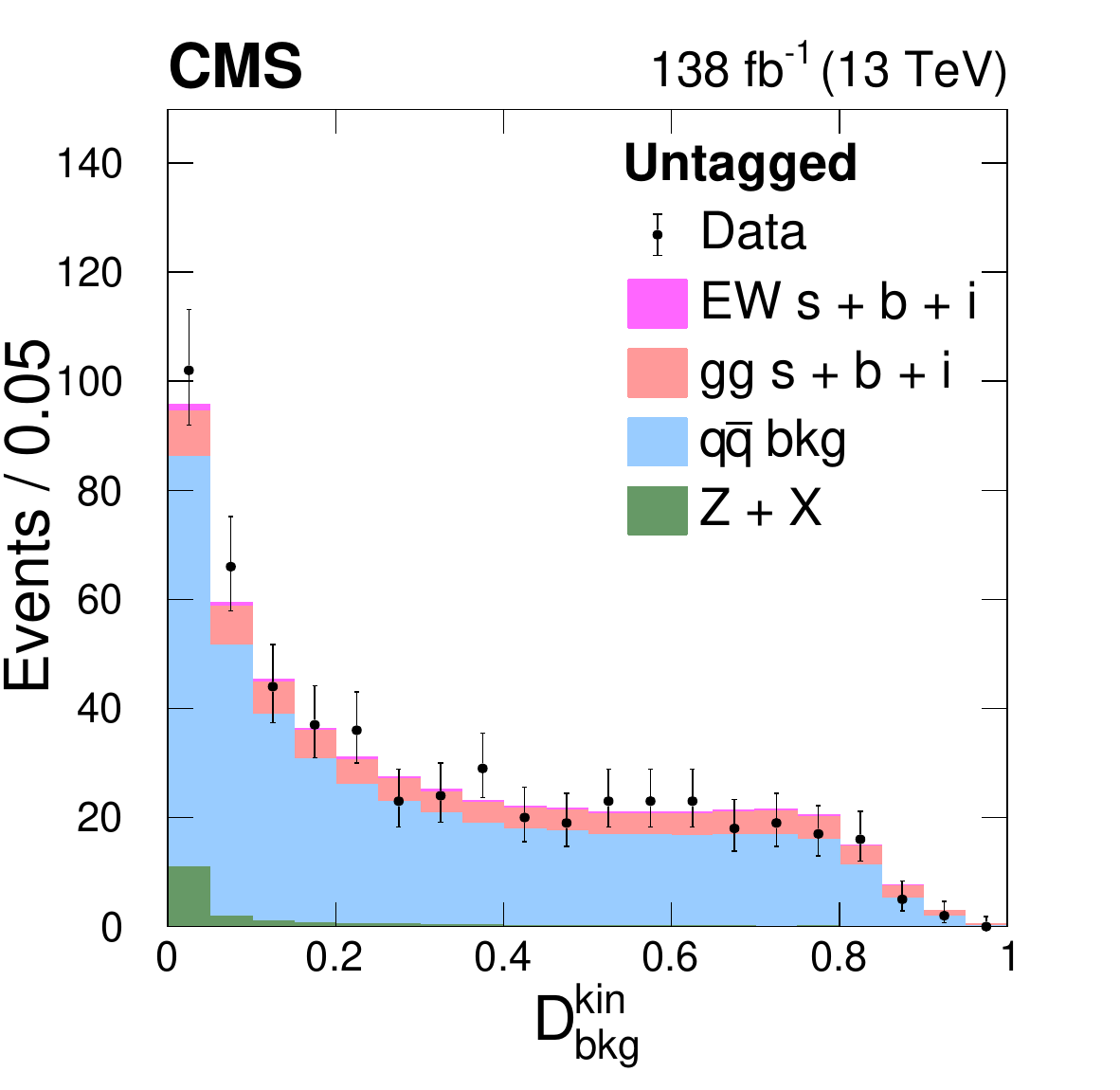}
\includegraphics[width=0.31\textwidth]{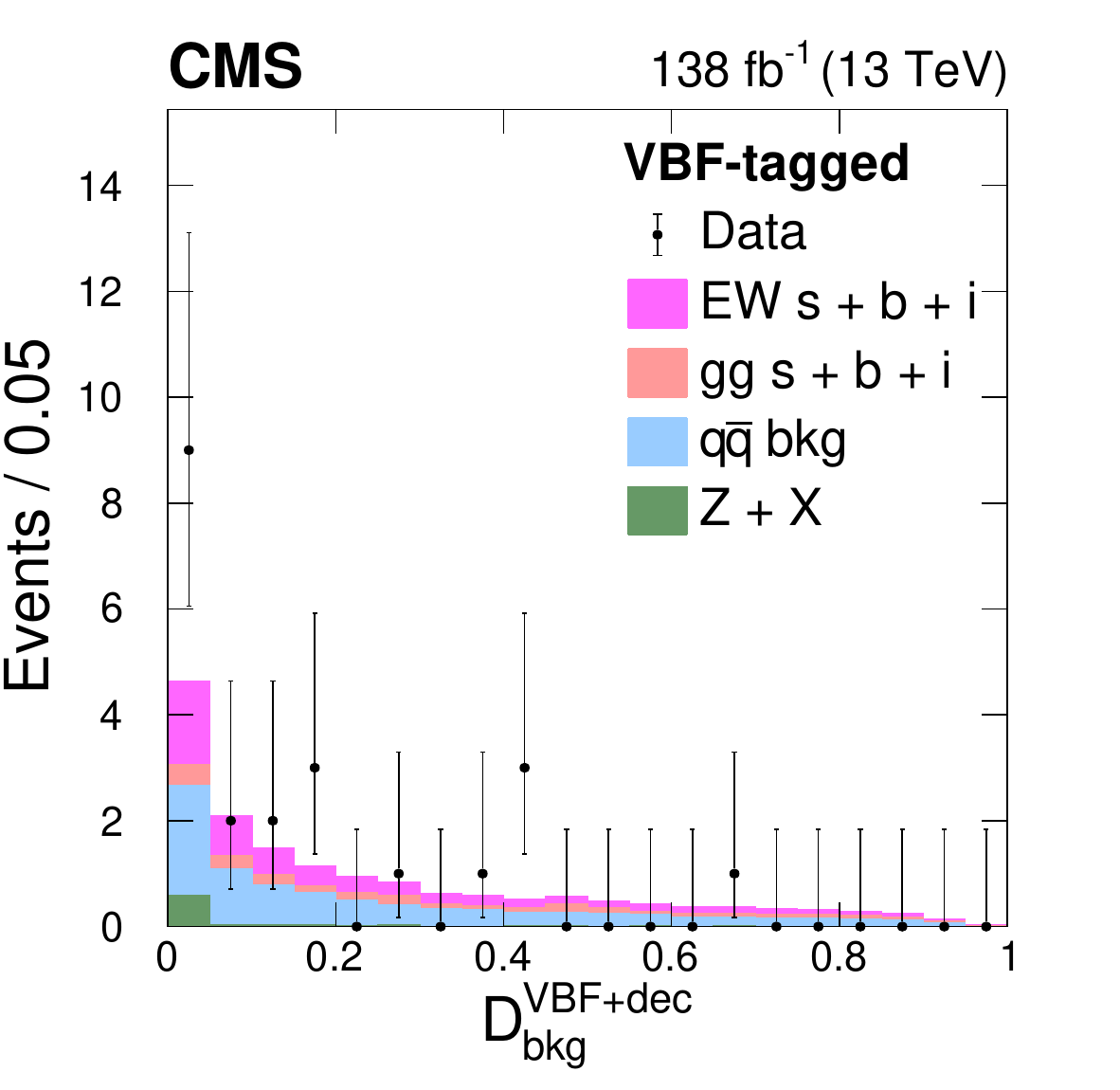} 
\includegraphics[width=0.31\textwidth]{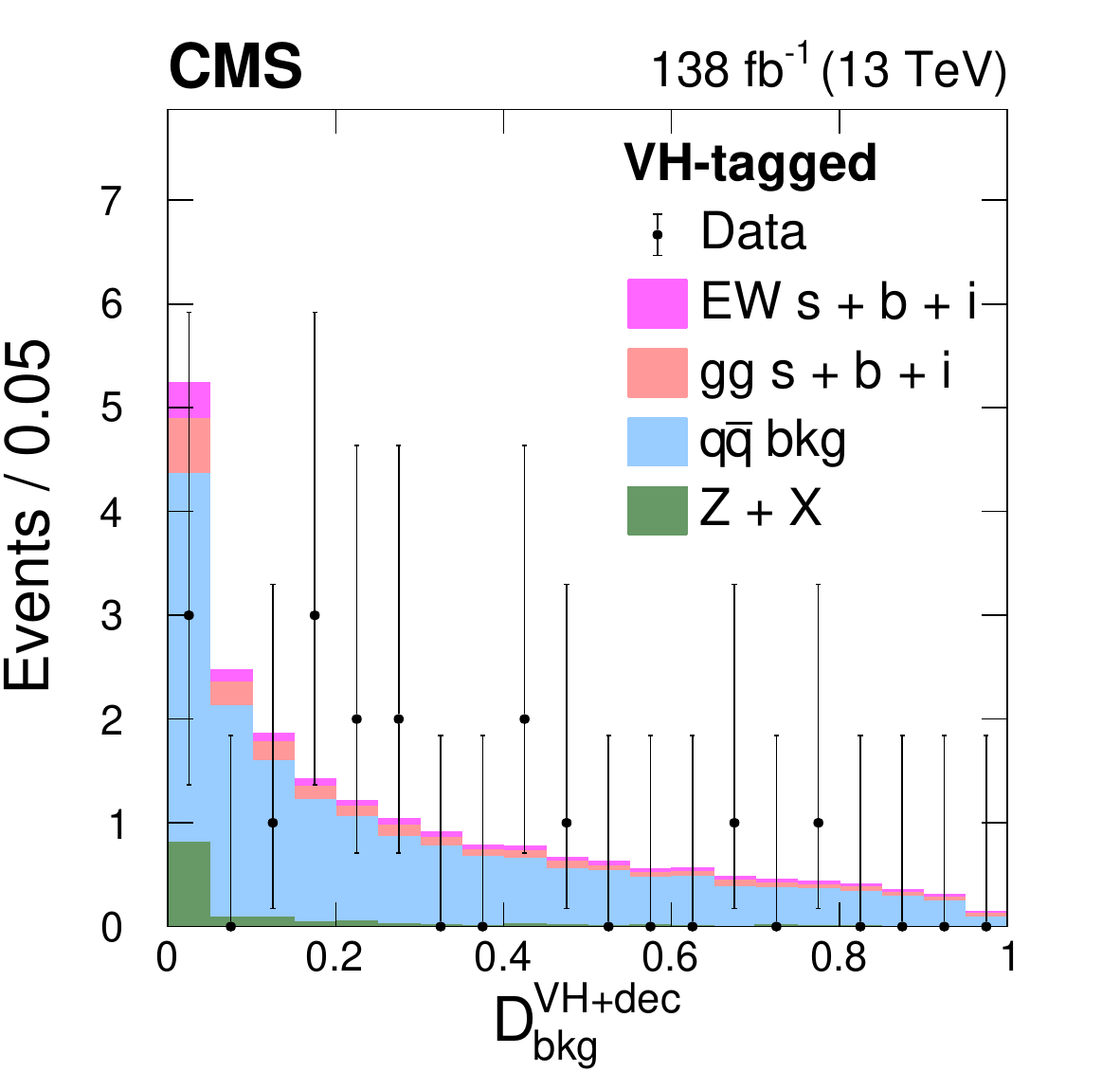}\\ 
\includegraphics[width=0.31\textwidth]{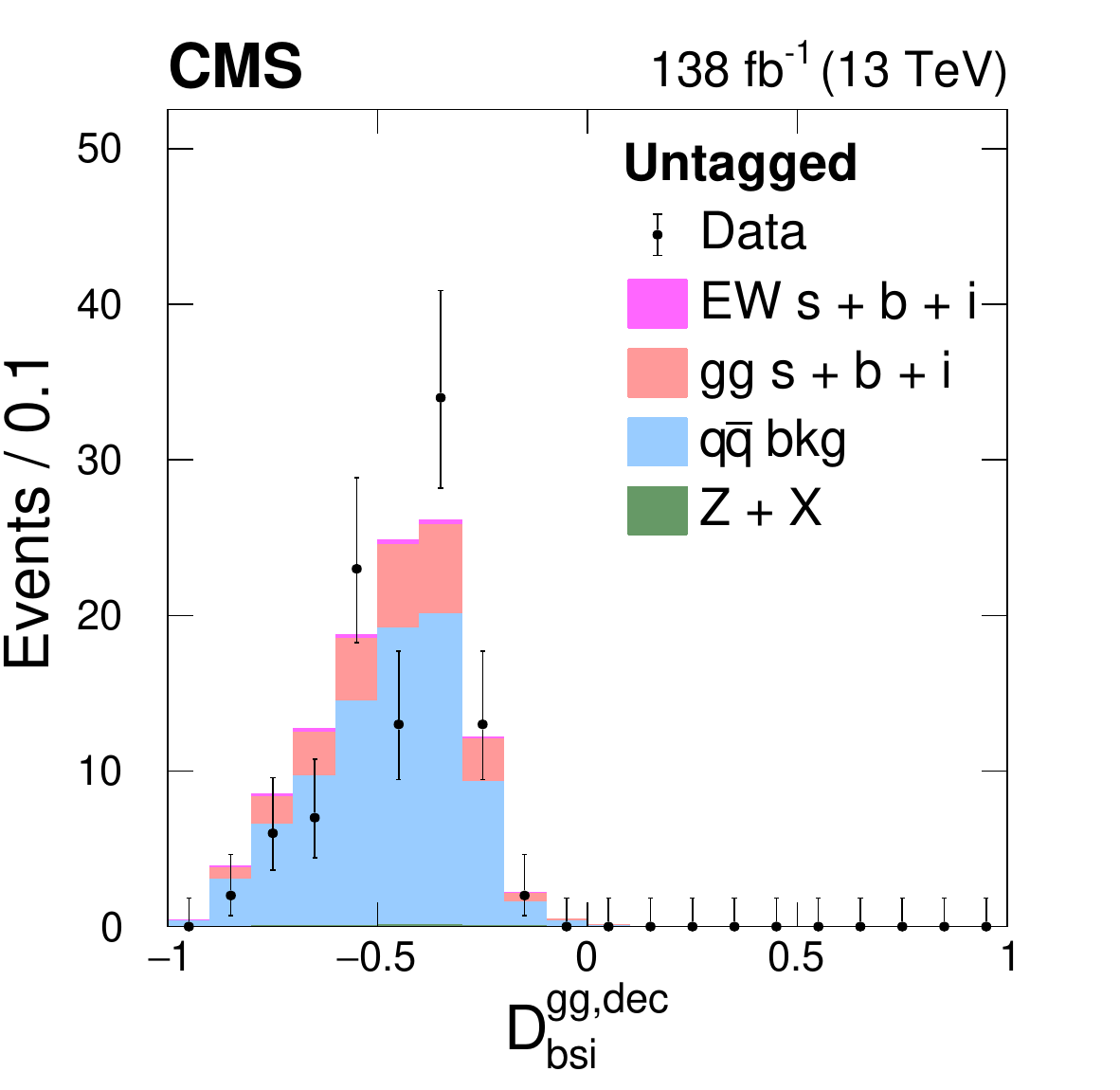}
\includegraphics[width=0.31\textwidth]{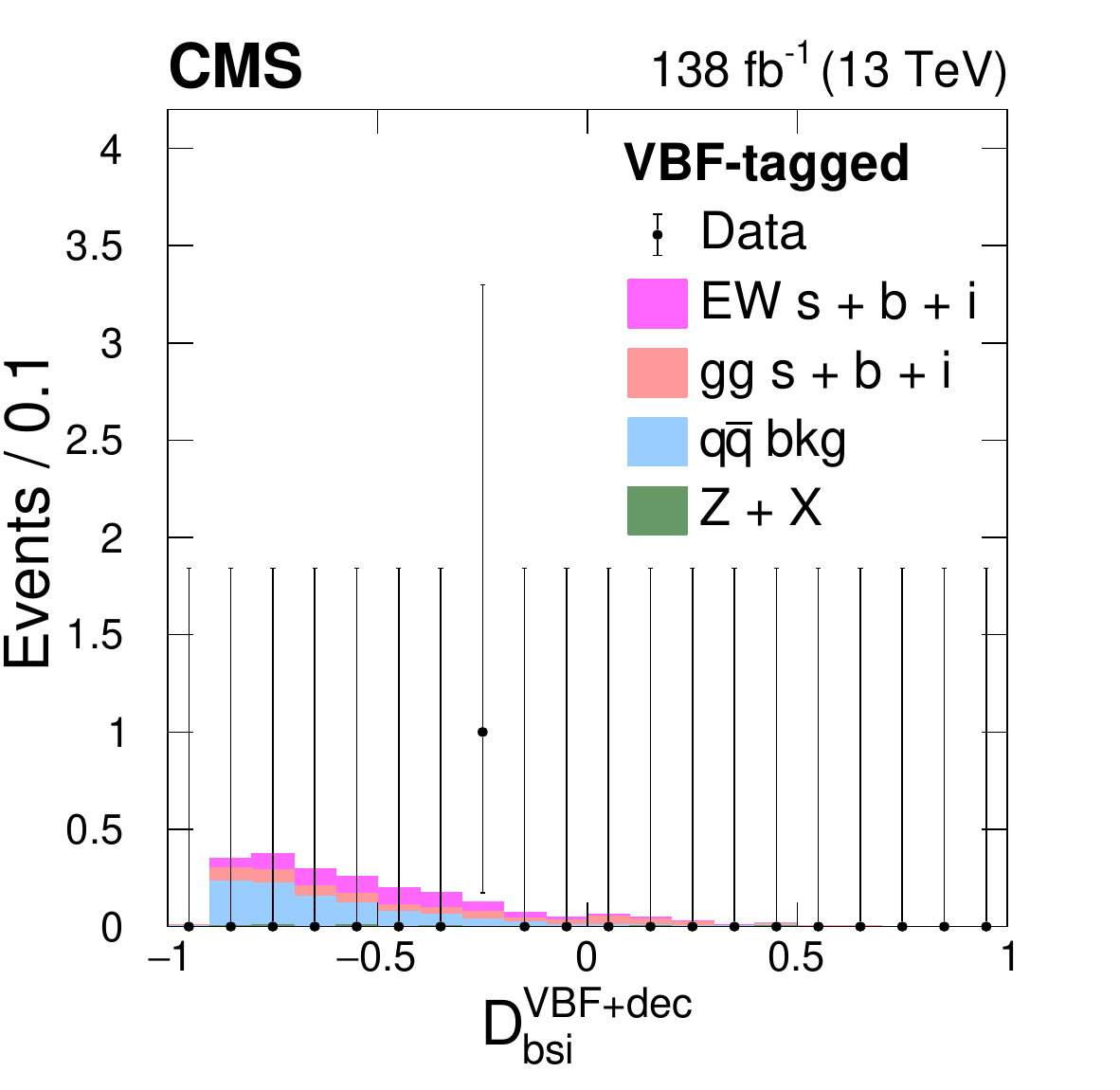} 
\includegraphics[width=0.31\textwidth]{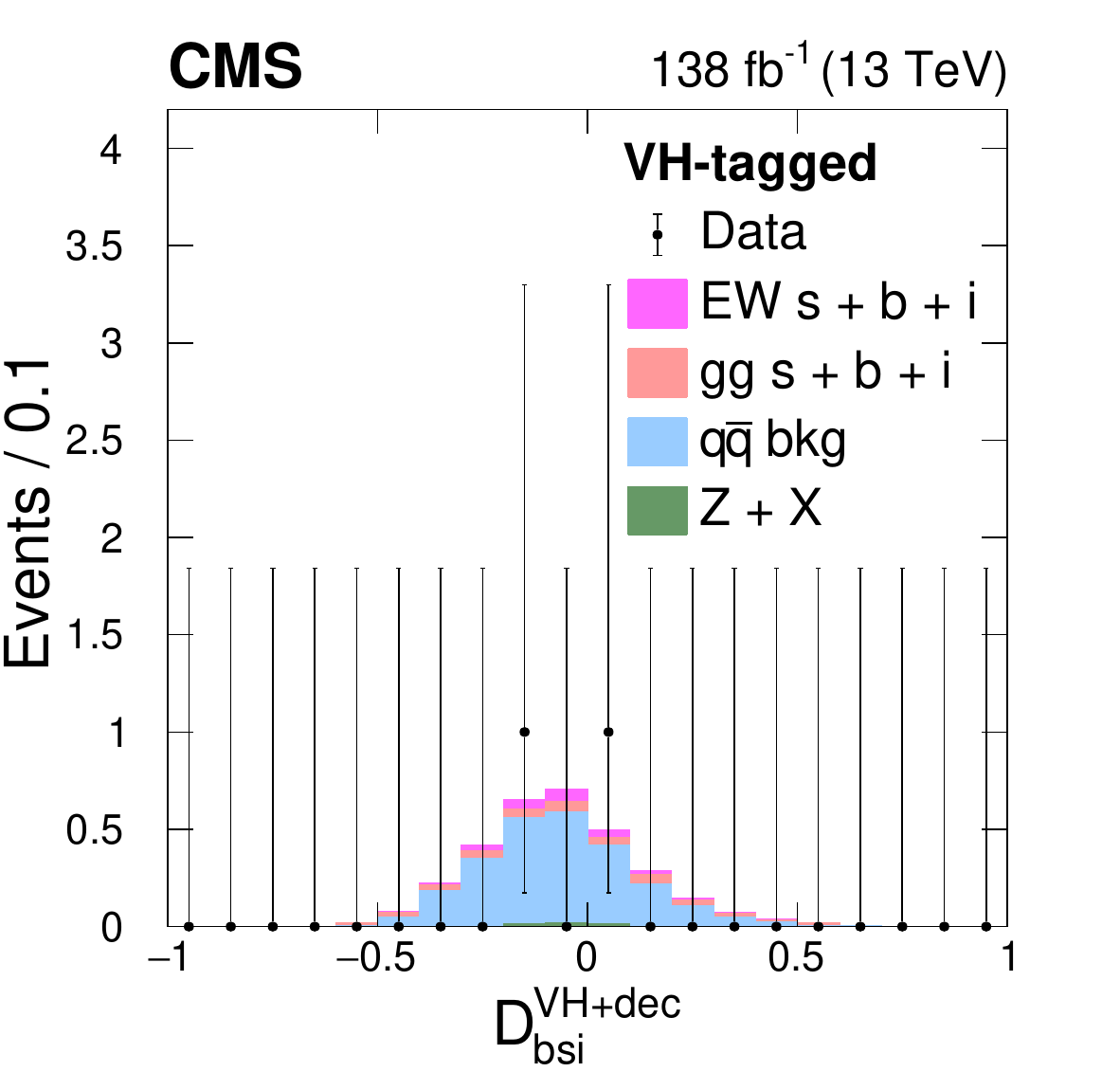}\\ 
\caption{
Off-shell data (points) and pre-fit distributions (histograms) for the Untagged (left), \VBF-tagged (middle), and \VH-tagged (right) categories.
The upper row shows $\mell$ distributions with a requirement on $\Dbkgkin>0.6$ (left), 
$\mathcal{D}^{{\VBF}+{\text{dec}}}_\text{bkg}>0.6$ (middle), or $\mathcal{D}^{{\VH}+{\text{dec}}}_\text{bkg}>0.6$ (right)
applied for illustration purposes to enhance signal over background contributions.
The middle row shows $\mathcal{D}^\text{kin}_\text{bkg}$ (left), $\mathcal{D}^{{\VBF}+{\text{dec}}}_\text{bkg}$ (middle), $\mathcal{D}^{{\VH}+{\text{dec}}}_\text{bkg}$ (right) distributions,
where an additional requirement $\mell>340$\GeV is applied to enhance signal-over-background contributions.
The lower row plots the $\mathcal{D}_\text{bsi}$ with both the \mell and \Dbkgkin requirements specified above.
Contributions from the four processes are shown by the different colors, where ``s'', ``b'', and ``i'' refer to the 
signal, background, and interference contributions, respectively.
The vertical bars on the points give the statistical uncertainties in the data, and the horizontal bars represent the bin widths. 
For the prefit distributions, the different cross sections are set to their SM values.
}
\label{fig:ObservablesCat}
\end{figure*}

In each category of events, three observables $\vec{x}$ are defined following Eqs.~(\ref{eq:melaD}) and~(\ref{eq:melaDint}).
As summarized in Table~\ref{tab:categoriesoffshell}, $\vec{x} = \{ \mell, \Dbkg, \Dbsi \}$.
The $\mell$ and $\Dbkg$ parameters have already been introduced earlier. 
The third observable, \Dbsi referred to as \Dint in Eq.~(\ref{eq:melaDint}), separates the interference
of \Hboson production (with SM-like couplings) and the background (used as the alternative model). 
In the Untagged category, decay information is used in the calculation of \Dbsi, as indicated with the label ``dec''. 
In the VBF- and VH-tagged categories, production information from the two associated jets is used, along with decay information.
The observed and expected distributions of observables $\vec{x}$ for events in the \offshell region are illustrated in Fig.~\ref{fig:ObservablesCat} 
for each of the three categories. The expected distributions are found using the SM predicted signal and background cross sections.

\section{Estimation of background}
\label{Bkg2}
The largest background to the \Hboson signal in the \Hell channel is from the process 
$\qqbar\to\ZZ/\PZ\PGg^*/\PGg^*\PGg^*\to 4\Pell$.
The $\Pg\Pg\to\ZZ/\PZ\PGg^*/\PGg^*\PGg^*\to 4\Pell$ background process, as well as 
the EW background, which includes the vector boson scattering and $\PV\PZ\PZ$ processes, 
interfere with off-shell \Hboson production and are discussed 
in more detail in Sections~\ref{sec:Bkg} and~\ref{SignalModelling}. 
The interference between \onshell \Hboson production and these backgrounds is, instead, negligible. 
In the \onshell region, the EW background includes other $\PV\PV\PV$, $\ttbar\PV\PV$, and $\ttbar\PV$ processes.
To model the $m_{4\Pell}$ distributions for each of these irreducible backgrounds, 
a Bernstein polynomial of third order is used in the $m_{4\Pell}$ range 105--140 \GeV.
 
An additional background to the \Hboson signal, referred to as $\PZ+\PX$ in the following, comes from processes in which decays 
of heavy-flavor hadrons, in-flight decays of light mesons within jets, or charged hadrons overlapping with \PGpz 
decays are misidentified as leptons. The main process contributing to this background is $\PZ$+jets production, which
is estimated from control regions in data. The control regions are defined by requiring a lepton pair, satisfying all 
the requirements of a ${\PZ}_{1}$ candidate, along with two additional opposite-sign 
leptons satisfying looser identification requirements than those in the main analysis. 
These four leptons are then required to pass the ${\PZ}_{1}, {\PZ}_{2}$ selection. The background yield in the signal region 
is obtained by weighting the control region events by the lepton misidentification probability ($f_{\Pell}$).
The $f_{\Pell}$ is defined as the probability of a non-prompt lepton to pass the analysis selection criteria.
A detailed description 
of the method can be found in Ref.~\cite{Sirunyan:2017exp}.
The shape of the $m_{4\Pell}$ distribution from $\PZ+\PX$ events is described by a Landau function.
Due to the low number of events, the functional form is extracted using a larger $m_{4\Pell}$ range of 70--770 \GeV.

The observed number of data events, expected background, and signal yields are listed 
in Tables~\ref{table:yields_SR} and \ref{tab:templateyields_offshell} for the on- and \offshell regions, respectively. 
The signal and $\PZ\PZ$ background yields are estimated from simulation, while the $\PZ+\PX$ yield is estimated from data.
The details of the \Hboson signal modeling, its interference with background, 
and the $\PZ\PH$ cross-feed for the \offshell results are given in Section~\ref{SignalModelling}. 

\begin{table*}[!htb]	
\centering
\topcaption{The observed and expected yields for the \Hboson signal and background contributions in the \onshell region $105<m_{4\Pell}<140\GeV$, for each of the four-lepton categories and the total.}
\label{table:yields_SR}
\begin{scotch}{lrrrrr}
   	&	4\PGm	&	4\Pe	&	$2\Pe2\PGm$	&	$2\PGm2\Pe$	& Total \\
\hline
Total signal    &       90.9    &       48.7    &       65.5    &       53.3    &       258.4   \\
$\qqbar\to4\Pell$ background 	&	89.2	&	38.9	&	64.4	&	42.1	&	234.6	\\
$\Pg\Pg\to4\Pell$ background	&	9.7	&	4.9	&	4.9	&	3.8	&	23.4	\\
$\PZ+\PX$ background	&	32.4	&	12.2	&	28.2	&	18.6	&	91.3	\\
Total expected	&	222.2	&	104.6	&	163.0	&	117.8	&	607.7	\\	
Observed	&	230	&	94	&	170	& 107	&	601	\\
\end{scotch}
\end{table*}

\begin{table*}[!htb]
\centering
\caption{Observed and expected yields for the \Hboson signal and background contributions 
in the \offshell region $m_{4\Pell}> 220\GeV$, for each of the four-lepton categories and the total. 
The yields from interference of the signal and background and the $\PZ\PH$ cross-feed are also shown. 
The expected yields are adjusted within their respective uncertainties from the fit to the data.}
\label{tab:templateyields_offshell}
\begin{scotch}{lrrrr}
& VBF-tagged & VH-tagged & Untagged & Total \\
\hline
$\Pg\Pg\to4\Pell$ signal &1.0 & 0.9 & 25.1 & 27.0 \\
$\Pg\Pg\to4\Pell$ background & 16.0 & 13.5 & 457.0& 486.4 \\
$\Pg\Pg\to4\Pell$ interference & $-2.1$ & $-1.9$  &  $-52.0$ & $-56.1$ \\
EW signal         & 1.3 & 0.1 & 1.8& 3.2\\
EW background     & 14.9 & 2.9 & 19.7& 37.5\\
EW interference     & $-3.4$ & $-0.2$ & $-3.9$ & $-7.4$ \\
$\PZ\PH$ cross-feed & 0.2 & 0.4 & 7.0& 7.6 \\
$\qqbar\to4\Pell$ background & 28.6 & 46.7 & 1795.1 & 1870.4  \\
$\PZ+\PX$ background & 4.7 & 5.3 & 89.7& 99.7  \\
Total expected & 61.2 & 67.8 & 2339.4 & 2468.4\\
Observed & 70 & 67 & 2335 & 2472 \\
\end{scotch}
\end{table*}

\section{Extraction of signal}
\label{SignalModelling}
The modeling of the signal process is different for the on- and \offshell regions. 
In the \onshell region, there is negligible interference between the \Hboson and background 
production, so the signal can be treated separately. At the same time, the narrow \Hboson
peak requires careful treatment of the detector resolution effects. 
There is also little dependence on the production processes because
the $\PH\to4\Pell$ decay is independent of the production mechanism 
due to the narrow-width approximation.
In the \offshell region, on the other hand, the distributions are much wider, 
with minimal influence from detector resolution on the measurements.
However, there is a sizable interference between the \Hboson signal 
and background processes, and their proper treatment requires special care. 
Moreover, the treatment of the gluon fusion and EW processes differs 
because of their different evolution with $m_{4\Pell}$.

\subsection{Signal modeling in the on-shell region}\label{onshellmodel}
The probability density for the on-shell region includes both signal and background contributions.  It is normalized to the total event yield for each process $j$ and category $k$, and can be written as
\begin{equation}
  \mathcal{P}_{jk}(\vec{x};\vec{\xi}_{jk},\vec\zeta) = 
  \mu_j\mathcal{P}_{jk}^\text{sig} (\vec{x};\vec{\xi}_{jk},\vec\zeta) +
  \mathcal{P}_{jk}^\text{bkg}(\vec{x};\vec{\xi}_{jk}),
  \label{eq:ponshell}
\end{equation}
where $\vec\zeta$ are the unconstrained parameters of interest,
including the signal strength $\mu_j$ defined as the ratio of the signal yield to the SM expectation,
$\vec{\xi}_{jk}$ are the constrained nuisance parameters for a particular parametrization,
and $\vec{x}$ are the observables.

In the \onshell \Hboson mass and width analysis, there are six signal processes (\ggH, \VBF, \VH, \ttH, \bbH, and \tqH) 
and three background processes ($\qqbar/\Pg\Pg\to\ZZ/\PZ\PGg^*/\PGg^*\PGg^*\to 4\Pell$ and $\PZ+\PX$).
When constraints on \GH are determined by simultaneously fitting both the \onshell and \offshell regions,
the \onshell region follows Scheme 2 as described in Ref.~\cite{CMS:2021nnc}, 
where additional details on categorization and signal and background modeling can be found.
Additional details regarding the signal modeling for the mass and 
width analyses using only the on-shell region are provided below.

\begin{figure}[!htb]
  \centering
  \includegraphics[width=0.45\textwidth]{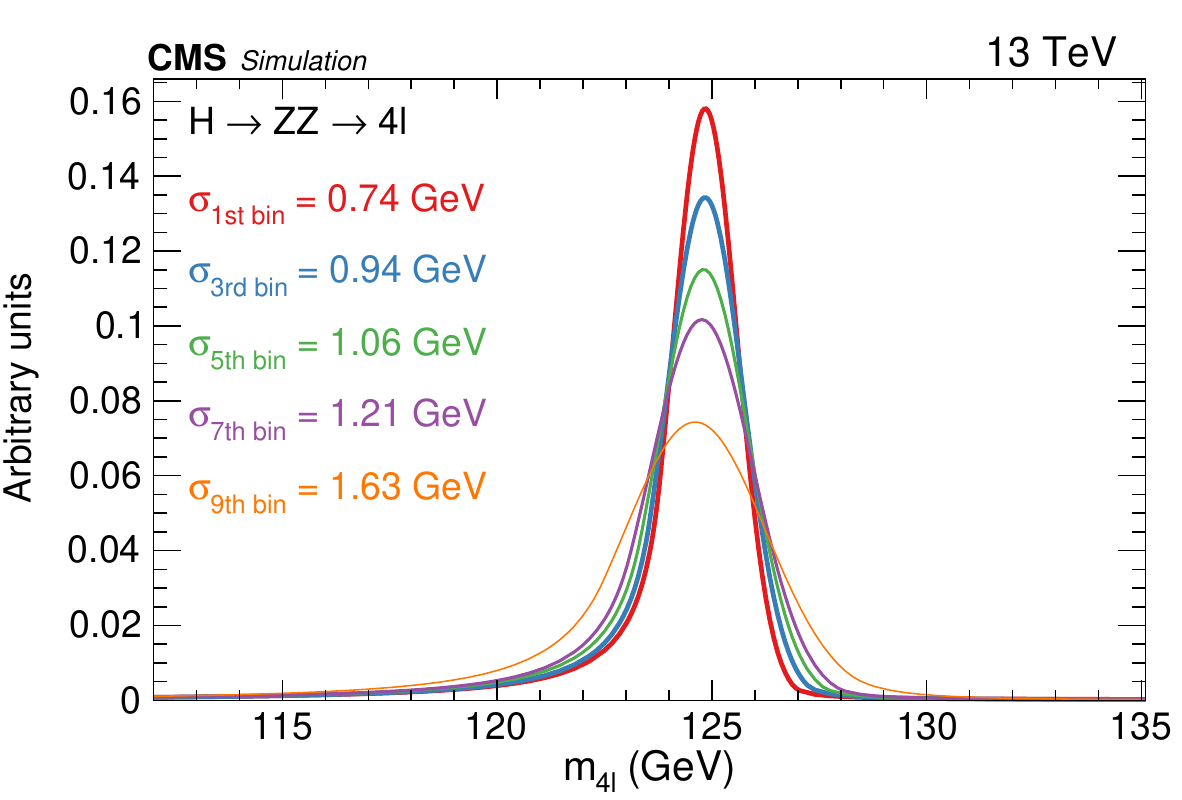}
  \caption{Examples of the \onshell signal model $m_{4\Pell}$ shape for five \RelMassErr bins. For illustration purposes, the mass range shown is reduced to 112--135 \GeV, and all final states and all data-taking years are combined. The resolution estimator $\sigma_{i~bin}$ is defined as the width of the Gaussian function from a fit to events in bin $i$ having the smallest mass window that contains 68\% of the signal events.}
  \label{1Dmodel_category}
\end{figure}

\begin{figure}[!htbp]
  \centering
\includegraphics[width=0.5\textwidth]{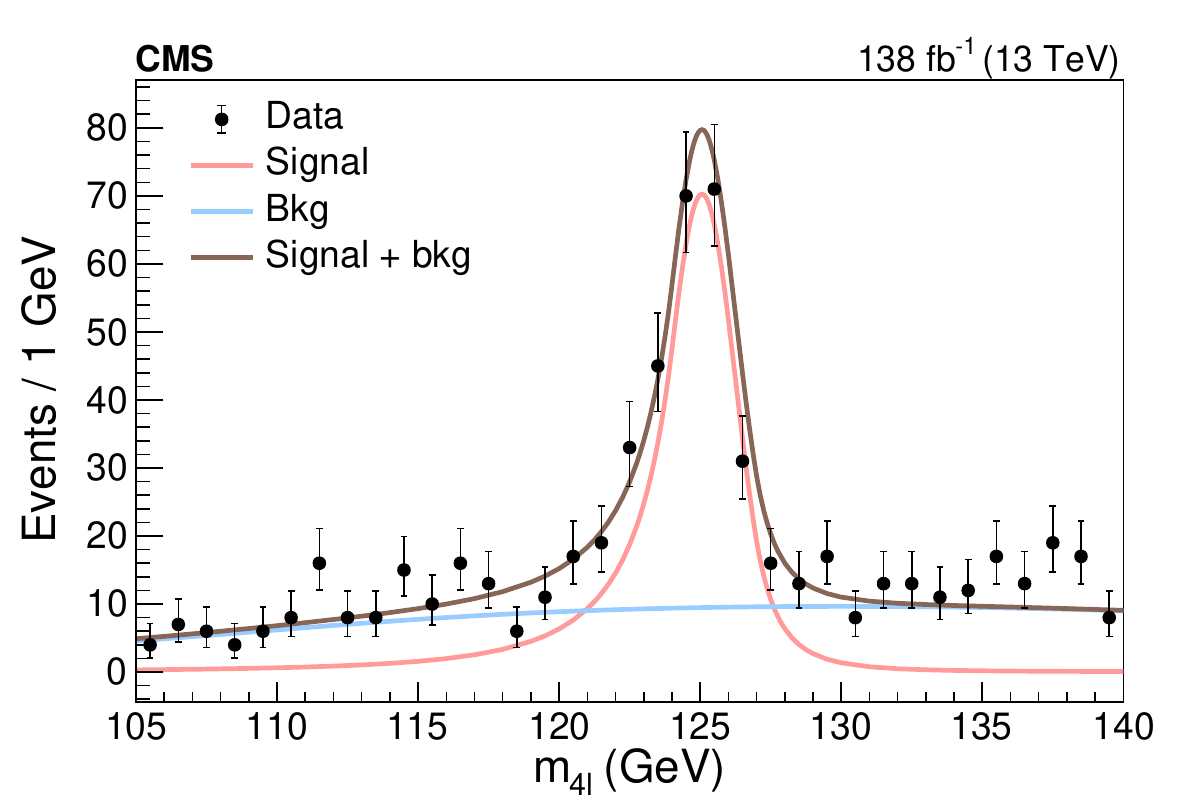}
	\caption{Illustration of how the on-shell statistical model is constructed, combining the $m_{4\Pell}$ distributions from all data-taking years and all final states. The red, blue, and brown lines show the results of the fit to the signal, background, and their sum, respectively. The solid black points with vertical bars show the data and the associated statistical uncertainties.}
\label{1D_model}
\end{figure}

For the \Hboson mass measurement, a one-dimensional (1D) statistical model is built in each category, per signal process, using the observable \mell, split 
between data-taking years and final states, and taking each of the nine \RelMassErr bins separately.
The signal shape for each process is obtained from a fit of the simulated \mell distribution, in the range 105--140\GeV, 
using a double-sided Crystal Ball function~\cite{Oreglia:1980cs}. In the case of \VH and \ttH production, a Landau function is added 
to include the possible contribution of a lepton from \Hboson decay being 
outside the detector acceptance or failing the selection criteria.
Five simulated mass points (120, 124, 125, 126, and 130 \GeV) are used to parametrize the \mH dependence of each shape parameter $p^{i}$ in the Crystal Ball and Landau functions. A set of first-order polynomials with constants $a^{i}$ and $b^{i}$ is fitted simultaneously as functions of \mH:
\begin{equation}
\label{eq:signalparams}
p^{i} = a^{i} + b^{i}(\mH - 125\GeV ).
\end{equation}
The natural width of the Higgs boson is assumed to be much smaller than the instrumental $m_{4\Pell}$ resolution. For the \onshell width measurement instead, the signal shape is the convolution of a Crystal Ball function with a Breit--Wigner function, to include \GH as a parameter in the model.
This is referred to as the 1D model. If the BS and \mZone constraints are applied, then it is referred to as 1D$^{'}_\text{BS}$ model.
Figure~\ref{1Dmodel_category} shows an example of the signal model used in the \onshell analysis and how it varies across the \RelMassErr bins. 
This categorization isolates the events with better invariant mass resolution and improves the description of the signal shape and the modeling of the peak position. 
This latter improvement is most important for the final state where the \onshell \PZ boson decays to two electrons. 
An illustration of the full model used in the \onshell analysis is shown in Fig.~\ref{1D_model}.

Based on the 1D model, a two-dimensional (2D) statistical model is built with a probability density $\mathcal{P}_\text{2D}$ using the observables \mell and \Dbkgkin and the expression 
\ifthenelse{\boolean{cms@external}}{
  \begin{multline}\label{eq:p2d}
    \mathcal{P}_\text{2D}(\mell,~\Dbkgkin) = \mathcal{P}^\text{sig}_\text{1D}~T^\text{sig}(\Dbkgkin\vert \mell)+\\
    \mathcal{P}^\text{bkg}_\text{1D}~T^\text{bkg}(\Dbkgkin\vert \mell).
  \end{multline}
}{
  \begin{equation*}\label{eq:p2d}
    \mathcal{P}_\text{2D}(\mell,~\Dbkgkin) = \mathcal{P}^\text{sig}_\text{1D}~T^\text{sig}(\Dbkgkin\vert \mell)+\mathcal{P}^\text{bkg}_\text{1D}~T^\text{bkg}(\Dbkgkin\vert \mell).
  \end{equation*} 
}

The factor $T$, based on a 2D histogram template, is a conditional probability density function of \Dbkgkin and \mell, 
implemented by normalizing the sum of the contents in each $m_{4\Pell}$ histogram bin to one.

The final model, designated $\mathcal{N}$--2D, is built by combining all 2D models of $\mathcal{N}$ different \RelMassErr categories. 
If the BS and \mZone constraint approaches are applied, the model is referred to as $\mathcal{N}$--2$D'_\text{BS}$. This configuration provides the best \Hboson mass and width precision and is used as the default for the \onshell measurements.
A simultaneous signal-plus-background unbinned fit to all categories is performed to extract \mH and \GH. The signal strength is left unconstrained in this fit. 

\subsection{Signal and background interference modeling in the off-shell region}

In the \offshell region, the probability density follows Eqs.~(\ref{eq:resonant}) and~(\ref{eq:ponshell}) closely,
with the additional contribution of the interference (``int'') between the signal and background amplitudes,
as well as a cross-feed (``cross'') component discussed below. It is parametrized as 
\ifthenelse{\boolean{cms@external}}{
  \begin{multline}\label{eq:poffshell}
    \mathcal{P}_{jk}(\vec{x};\vec{\xi}_{jk},\vec\zeta) =
    \frac{\mu_j \GH}{\Gref}\mathcal{P}_{jk}^\text{sig} ( \vec{x};\vec{\xi}_{jk})
    + \sqrt{\frac{\mu_j \GH}{\Gref}}\mathcal{P}_{jk}^\text{int} ( \vec{x};\vec{\xi}_{jk}) \\
    + \mu_j\mathcal{P}_{jk}^\text{cross} ( \vec{x};\vec{\xi}_{jk})
    + \mathcal{P}_{jk}^\text{bkg} ( \vec{x};\vec{\xi}_{jk}),
  \end{multline}
}{
  \begin{equation}\label{eq:poffshell}
    \mathcal{P}_{jk}(\vec{x};\vec{\xi}_{jk},\vec\zeta) =
    \frac{\mu_j \GH}{\Gref}\mathcal{P}_{jk}^\text{sig} ( \vec{x};\vec{\xi}_{jk})
    + \sqrt{\frac{\mu_j \GH}{\Gref}}\mathcal{P}_{jk}^\mathrm{int} ( \vec{x};\vec{\xi}_{jk})
    + \mu_j\mathcal{P}_{jk}^\text{cross} (\vec{x};\vec{\xi}_{jk})
    + \mathcal{P}_{jk}^\text{bkg} ( \vec{x};\vec{\xi}_{jk}),
  \end{equation}
}
where $\Gref$ is the reference value of the \Hboson width (not necessarily its SM value)
used in simulation. Otherwise, the definition of the terms is the same as in Eq.~(\ref{eq:ponshell}),
including $\mu_j$, which represents the signal strength in the \onshell region.
In the \offshell width analysis, there are two production modes, $j$ (gluon fusion and EW, which includes both \VBF and \VH), 
and three jet-tagged categories, $k$. All lepton flavors and data periods are combined in this analysis.
The contributions from \ttH, \bbH, and \tqH are expected to be negligible in the \offshell region.

The $\mathcal{P}_{jk}^{\text{sig}}$, $\mathcal{P}_{jk}^{\text{int}}$, $\mathcal{P}_{jk}^{\text{cross}}$, and $\mathcal{P}_{jk}^\text{bkg}$ probability densities
are normalized to the expected number of events, and are implemented as binned histograms (templates)
of the observables $\vec{x}$ listed in Table~\ref{tab:categoriesoffshell}.
These templates are obtained as weighted linear combinations of existing simulated signal or background samples.

The \offshell region includes all events with $m_{4\Pell}$ greater than $220\GeV$. 
Other processes can mimic \offshell \Hboson production and decay to four leptons in this region. 
Specifically, we study \onshell signal events in which the \Hboson decays 
to $2\Pell +X$, and $X$ contains further leptons that allow the event to pass the four-lepton selections. 
The dominant \onshell \Hboson process that can contaminate the \offshell region is 
$\PZ(\Pell\Pell)\PH(2\Pell+X)$, called $\ZH$ cross-feed. 

The $\ZH$ cross-feed contribution is estimated using \onshell 
$\ZH$ samples with $\PH \to \WW$ and $\ZZ$ decays, 
where both hadronic and leptonic decay of the \PZ bosons is allowed. 
The dominant contributions come from the $\PH \to 2\Pell 2\PGn$ and $2\Pell 2\Pq$ final states
produced in association with $\PZ\to 2\Pell$. 
To prevent double counting, \onshell cross-feed events are eliminated from the \offshell simulated samples.

The gluon fusion cross section is calculated using the highest order QCD and EW expansions available 
to simulate inclusive $\PZ\PZ$ production~\cite{deFlorian:2016spz}.
However, event categorization depends on modeling associated jets through \PYTHIA's parton showering and hadronization, 
which involves matching these processes to the hard-scattering production. Off-shell gluon fusion production is generated at LO 
with no associated jets at the matrix element level, and therefore all jets come from \PYTHIA. 
The parton showering and hadronization requires setting the hadronization scale, which can depend on the energy scale 
in the process. In the case of the $\Pg\Pg\to 4\Pell$ process, the energy scale is set at $m_{4\Pell}$.

The EW cross section for the inclusive production of $\PZ\PZ$ and two associated jets is also calculated to the highest order 
QCD and EW expansion available~\cite{deFlorian:2016spz}. 
Contrary to the gluon fusion process, two hard jets, which are typically the leading 
jets in the process, are already generated at the matrix element level in the LO simulation. 
These are the two associated jets in the VBF process, or 
the two jets from the hadronic decay of the associated \PW or \PZ boson. 
Therefore, the dependence on the \PYTHIA parton shower and hadronization 
and its matching to the hard-scattering production is much weaker for these EW processes. 

The categorization efficiencies, defined as the fractions of events distributed across the three categories, 
of simulated \ggH and EW \Hboson production can be checked using alternative \POWHEG and \minlo simulations. 
The \POWHEG samples are generated for a wide range of \offshell \Hboson masses at NLO in QCD, 
with one jet generated at the matrix element level, and using \PYTHIA matching to simulate additional jets. 
The \minlo simulation of \ggH production is done for \Hboson masses of 125 and $300\GeV$ at NLO in QCD, 
with two jets generated at the matrix element level, and \PYTHIA matching for additional jets.
While the \jhugen categorization efficiencies agree with those using \POWHEG and \minlo 
within the uncertainties of the QCD scale used in \PYTHIA,
for the \ggH process the deviations of the central values and the corresponding uncertainties 
are up to 20\% in the signal-dominated $m_{4\Pell}$ range 300--500\GeV, depending on the category. 
In the EW process, categorization efficiencies from the two approaches typically agree within 5--10\%. 
In all cases, we adjust the categorization efficiency as a function of $m_{4\Pell}$ to match that for the SM signal 
obtained from the \POWHEG samples, and assume the same behavior for 
the background and interference contributions. 
This correction procedure ensures that the total event yield, summed over the three categories,
is unchanged at each value of $m_{4\Pell}$.

Simulation of the $\vec{x}$ observables listed in Table~\ref{tab:categoriesoffshell}
is not affected by the jet modeling in the Untagged category. It is also found that the modeling
of the observables in the jet-tagged categories is nearly the same in the EW process, when 
compared between the direct \MCFM + \jhugen samples and reweighted \POWHEG + \jhugen. 
However, the modeling of jets in the jet-tagged categories for the \ggH process does impact 
the parametrization of the probability distributions. Therefore, within these two jet-tagged categories, 
the gluon fusion process is incorporated through the reweighted \POWHEG + \jhugen simulation. 
This approach allows a more precise matching with the parton shower, thereby 
better describing the associated jet activity. In this case, the samples are reweighted for the appropriate 
model using the \mela package, as discussed in Section~\ref{sec:Bkg}.

\section{Systematic uncertainties}
\label{sec:systUncert}

Several systematic uncertainties are estimated in the measurement of the constrained parameters $\vec{\xi}_{jk}$.
The template shapes describing the probability distributions in Eqs.~(\ref{eq:ponshell}) and~(\ref{eq:poffshell}) 
are varied separately within either the theoretical or experimental uncertainties, and the resulting variation in the constrained parameters is taken as the systematic uncertainty from this source.

The largest systematic uncertainties in the \onshell determination of the \Hboson mass and width are in the lepton momentum scale and resolution. 
The estimates of these uncertainties are extracted by combining two effects. First, the systematic uncertainties affecting the corrections are propagated to the Higgs boson events. 
Relative peak position shift and width changes, with respect to the nominal distribution, are considered as systematic uncertainties of the model. 
Then, since $2\ell$ events used to extract calibrations have a different momentum distribution with respect to the \Hboson ones, a closure test comparing data and simulation in $\PZ \to 2\ell$ events, as a function of lepton \pt and $\abs{\eta}$ is produced.
The differences observed at this stage are propagated to the \Hboson events and relative deviations from the nominal peak position and width values are considered as additional sources of systematic uncertainty. 
Considering both effects, the estimated systematic uncertainties are 0.03 and 0.15\% in the muon and electron momentum scales, respectively, and 3 and 10\% for resolution.

The largest systematic uncertainty in the \offshell measurement of the \Hboson width is from the modeling
of the dominant background process $\qqbar\to\ZZ/\PZ\PGg^*/\PGg^*\PGg^*\to 4\Pell$. 
Experimental uncertainties specific to the \offshell analysis involve those from the jet energy calibration, 
which are only relevant for the \VBF- and \VH-tagged categories.

Theoretical uncertainties that affect both the signal and background estimations include those from 
the renormalization and factorization scales, and the choice of the parton distribution function set. The uncertainty from the renormalization 
and factorization scales is determined by varying these scales
by factors of 0.5 and 2 from their nominal values while keeping the ratio of the two scales between 0.5 and 2.
The uncertainty due to the parton distribution function set is estimated by taking the root mean square of the variations
when using different replicas of the default NNPDF set. An additional uncertainty of 10\% in the $K$~factor used for the \ggZZ prediction is applied.

The integrated luminosities for the 2016, 2017, and 2018 data-taking years have 1.2--2.5\% 
individual systematic uncertainties~\cite{CMS-LUM-17-003,CMS-PAS-LUM-17-004,CMS-PAS-LUM-18-002}, 
while the overall uncertainty for the 2016--2018 period is 1.6\%. 
The uncertainty in the lepton identification, reconstruction, and selection efficiency ranges 
from 2 to 14\% in terms of the overall event yield for the 4$\PGm$ and 4\Pe final states, 
respectively, and affects both signal and background processes. 

In the estimation of the \ZX background, the flavor composition of hadronic jets misidentified as leptons can
be different in the $\PZ+1\Pell$ and $\PZ+2\Pell$ control regions. Together with the statistical uncertainty
in the $\PZ+2\Pell$ region, this uncertainty accounts for about a 30\% variation in the background yields.
The uncertainty in the modeling of this misidentification rate as a function of \pt and $\eta$,
combined with the $\PZ+1\Pell$ control region statistical uncertainty, leads to
uncertainties in the backgrounds yields ranging from 32\% in the 4\Pe final state to 39\% in the 4$\PGm$.

\section{Higgs boson mass and width measurements with on-shell production}
\begin{figure}[!htb]
  \centering
  \includegraphics[width=0.45\textwidth]{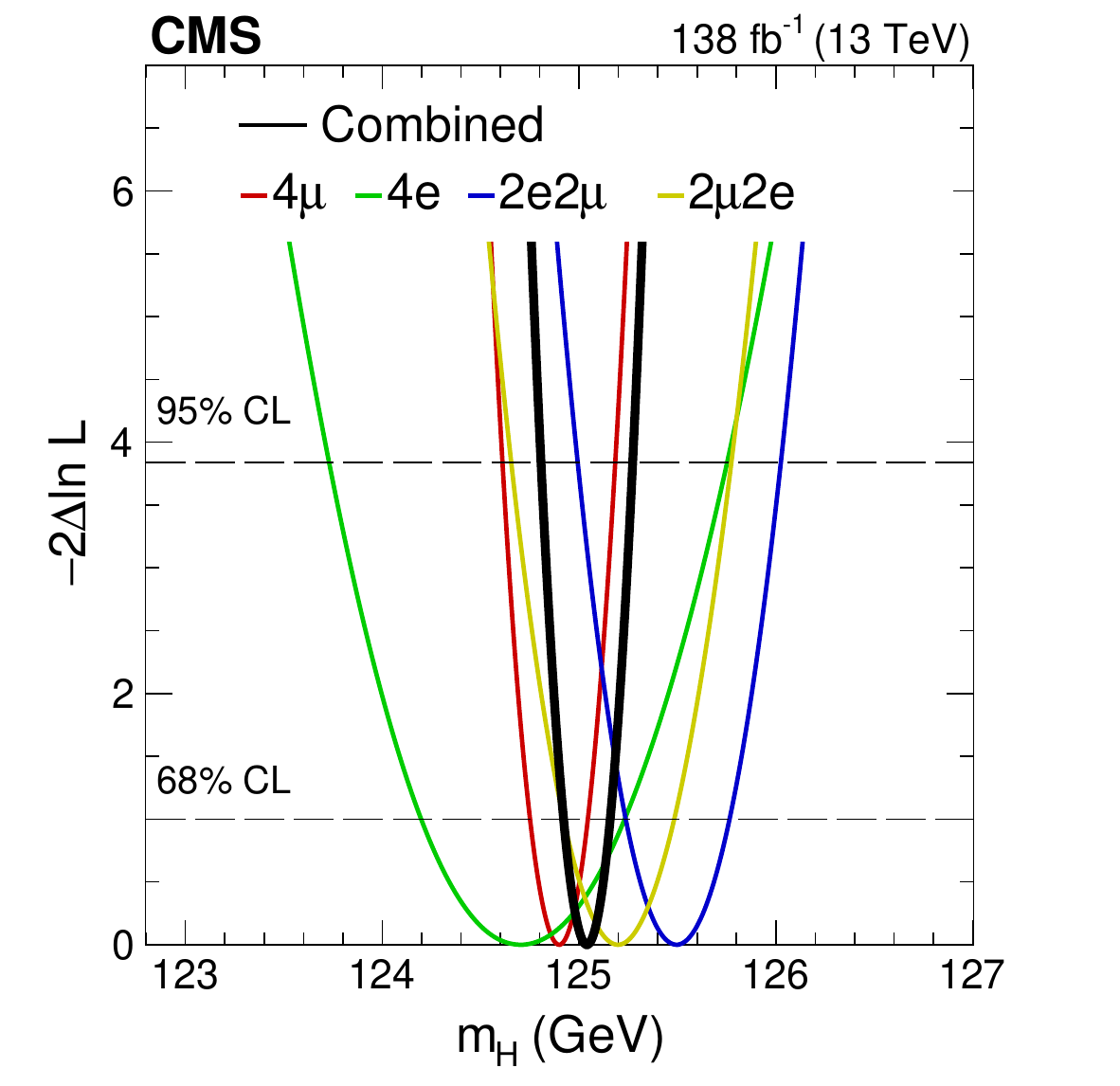}
  \caption{The profile likelihood from the \mH fit using the $\mathcal{N}$--2D$'_\text{BS}$ model for each of the 4$\Pell$ categories and combined.  The change in likelihood corresponding to 68 and 95\% CLs are shown by the dashed horizontal lines. Both statistical and systematic uncertainties are included in the fits.}
  \label{MassLikelihoodScan}
\end{figure}

The \Hboson mass and width are measured, using \onshell production, by fitting the \mell distribution in the mass range $105< \mell < 140\GeV$, using different likelihood models as described in Section~\ref{onshellmodel}.
The results have been determined using the CMS statistical analysis tool \textsc{Combine}~\cite{CMS:2024onh}, 
which is based on the \textsc{RooFit}~\cite{Verkerke:2003ir} and \textsc{RooStats}~\cite{Moneta:2010pm} frameworks.\\
Table~\ref{table:Mass1Dresults} shows
the mass measurements obtained from the 1D approach, where no further assumptions have been made.
In comparison to the 1D model, the 1D$^{'}_\text{BS}$ model (as described in Section~\ref{sec:LepImprov}) reduces the uncertainty by about 15\%.
Implementing the \RelMassErr categorization then gives the $\mathcal{N}$--1D$'_\text{BS}$ model, which leads to an additional 10\% improvement. 
Finally, using the \Dkinbkg discriminant to reduce the background produces the $\mathcal{N}$--2D$'_\text{BS}$ model with another 4\% improvement. 
Table~\ref{table:Mass_results} shows the resulting $m_{4\Pell}$ measurements using this last model. 
All the measured $m_{4\Pell}$ values from the different fits are statistically compatible, given their uncertainties and correlations.
Figure~\ref{MassLikelihoodScan} displays the observed 1D likelihood scans as functions of \mH, from the fits for the different $4\Pell$ categories and combined.
Combining all the $m_{4\Pell}$ final states and data-taking years, our final result is 
$m_\PH = 125.04~\pm~0.11$ (stat)$~\pm~0.05$ (syst) = $125.04~\pm~0.12$\GeV.
The largest systematic uncertainty is from the lepton momentum scale and equals 0.03 and 0.04 \GeV for final states with muons and electrons, respectively.

\begin{table*}[!htb]	
  \centering
  \topcaption{
    Best fit values for the mass of the \Hboson measured in the inclusive 4$\Pell$ final state and separately for different flavor categories using the 1D approach.
    Uncertainties are separated into statistical and systematic uncertainties.
    Expected uncertainties are also given assuming $m_{\PH} = 125.38\GeV$ \cite{Sirunyan:2020xwk}.
  }
  \renewcommand{\arraystretch}{1.6}
  \cmsTable{
    \begin{scotch}{lcc}
      $4\Pell$ category & Observed ($\pm$stat$\pm$syst) ($\GeVns$) & Expected uncertainty ($\pm$stat$\pm$syst)  ($\GeVns$) \\
      \hline
      Inclusive &	$124.98 \pm 0.14 \pm0.05$ & $ \pm 0.14 \pm 0.05$ \\
      $4\PGm$ &	$124.87 \pm 0.17\pm0.04$ & $\pm 0.18 \pm 0.04$ \\
      $2\PGm2\Pe$ &	$125.32^{+0.36}_{-0.37}\pm0.10$ & $^{+0.34+0.09}_{-0.34-0.10}$ \\
      $2\Pe2\PGm$ &        $125.23^{+0.36+0.09}_{-0.35-0.11}$ & $^{+0.35+0.09}_{-0.35-0.10}$ \\
      ${4\Pe}$ &    $124.94^{+0.68}_{-0.71} \pm 0.20$ & $^{+0.51+0.18}_{-0.51-0.20}$ \\
  \end{scotch}}
  \label{table:Mass1Dresults}
\end{table*}

\begin{table*}[!htb]	
  \centering
  \topcaption{Best fit values for the mass of the \Hboson measured in the inclusive 4$\Pell$ final state and separately for different flavor categories, using the final fit configuration ($\mathcal{N}$--2D'$_\text{BS}$).  Uncertainties are separated into statistical and systematic uncertainties. Expected uncertainties are also given assuming $m_{\PH} = 125.38\GeV$ \cite{Sirunyan:2020xwk}.}
  \renewcommand{\arraystretch}{1.6}
  \cmsTable{
    \begin{scotch}{lccc}
      $4\Pell$ category & Observed ($\pm$stat$\pm$syst) ($\GeVns$) & Expected uncertainty ($\pm$stat$\pm$syst) ($\GeVns$)  \\
      \hline
      Inclusive  & $125.04\pm0.11 \pm 0.05$	&	 $\pm 0.11\pm  0.05$ \\
      $4\PGm$ & $124.90 \pm 0.14 \pm 0.05$	&	$\pm 0.14\pm  0.04$ \\
      $2\Pe2\PGm$ &        $125.50^{+0.25}_{-0.24}\pm0.10$      &       $\pm0.24\pm0.10$ \\
      $2\PGm2\Pe$ &        $125.20^{+0.27+0.11}_{-0.26-0.07}$     &        $\pm0.27\pm0.10$ \\
      $4\Pe$ &	$124.70^{+0.49}_{-0.47}\pm0.20$     &	$\pm0.38\pm0.20$ \\ 
  \end{scotch}}
  \label{table:Mass_results}
\end{table*}

As a check on the analysis technique and the systematic uncertainty from this method, the 
1D$^{'}_\text{BS}$ model is applied to $\PZ \to 4\Pell$ events in the $m_{4\Pell}$ range 70--105 \GeV. The signal shape is obtained using a convolution of a Breit--Wigner function and a double-sided Crystal Ball function.
The fitted values of \mZ in different subchannels are $m_\PZ^{4\PGm} = 91.02 \pm 0.14\GeV$, $m_\PZ^{4\Pe} = 91.18 \pm 0.45\GeV$, $m_\PZ^{2\Pe2\PGm} = 91.40 \pm 0.29\GeV$, and $m_\PZ^{2\PGm2\Pe} = 91.40 \pm 0.37\GeV$, leading to a combined value of $m_\PZ = 91.17 \pm 0.12\GeV$, consistent with the world-average \PZ boson mass~\cite{Agashe:2014kda} and with the uncertainty in agreement with the expected value of $\pm~0.12$ \GeV from simulation.

\begin{figure*}[!htb]
  \centering
  \includegraphics[width=0.60\textwidth]{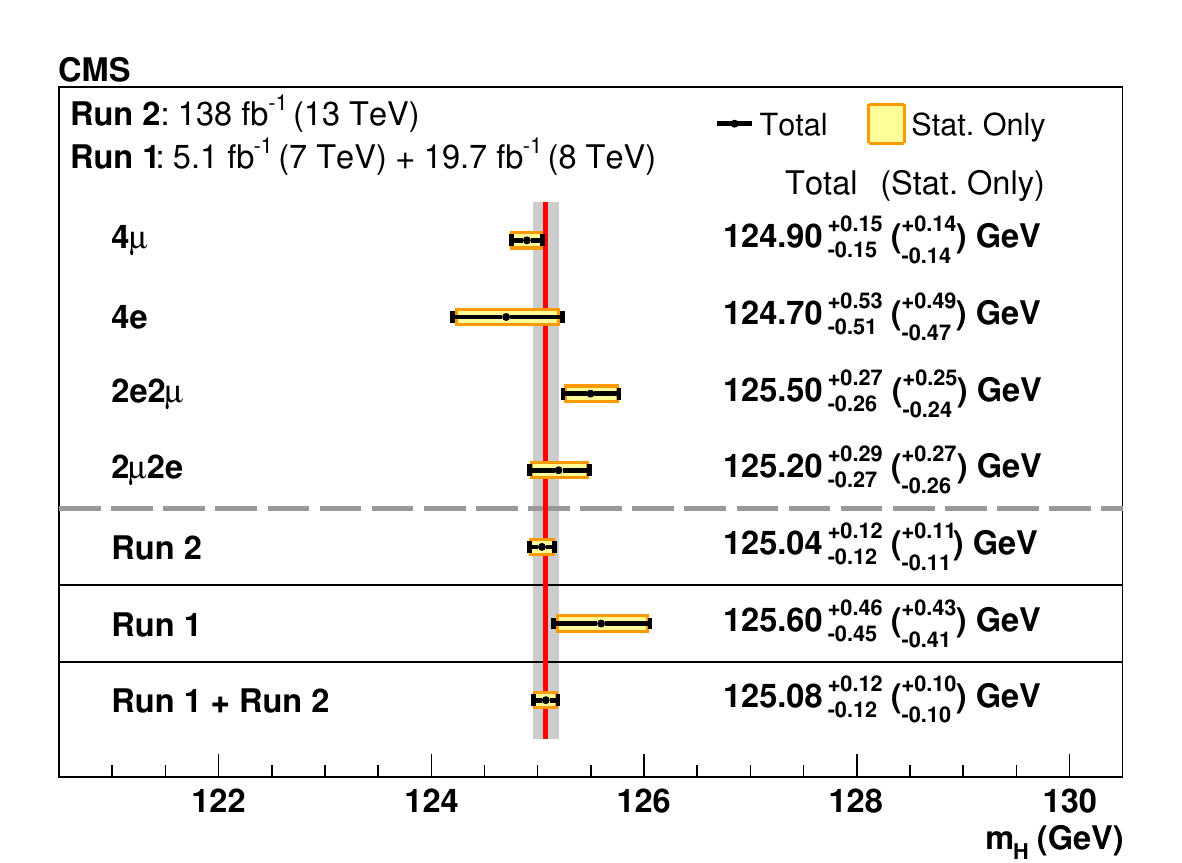}
  \caption{Summary of the CMS Higgs boson mass measurements using the four-lepton final state. The red vertical line and the gray column represent the best fit value and the total uncertainty, respectively, as measured by combining the Runs 1 and 2 data. The yellow band and horizontal black bars show the statistical and total uncertainties in each measurement, respectively. The value of each measurement is given, along with the total and statistical only (in parentheses) uncertainties.}
  \label{Run1Run2_scan}
\end{figure*}

The results from this analysis are combined with those extracted using data recorded with the  
CMS detector during Run 1 at $\sqrt{s}= 7$ and 8\TeV~\cite{Chatrchyan:2013mxa}. 
Since this analysis uses an improved method to extract the systematic uncertainties affecting lepton momentum, 
the lepton energy scales and resolution uncertainties are considered uncorrelated between the two runs. The combined observed result from
both data-taking periods is $\mH = 125.08 \pm 0.12\GeV = 125.08 \pm 0.10\stat \pm0.05\syst$\GeV.
The corresponding expected statistical and systematic uncertainties are $\pm$0.10 and $\pm$0.05\GeV, respectively. 
Figure~\ref{Run1Run2_scan} presents a summary of the \Hboson mass measurements by the CMS Collaboration in the four-lepton decay channel.

\subsection{Higgs boson width measurement from \onshell production}
\begin{figure}[!htb]
  \centering
  \includegraphics[width=0.5\textwidth]{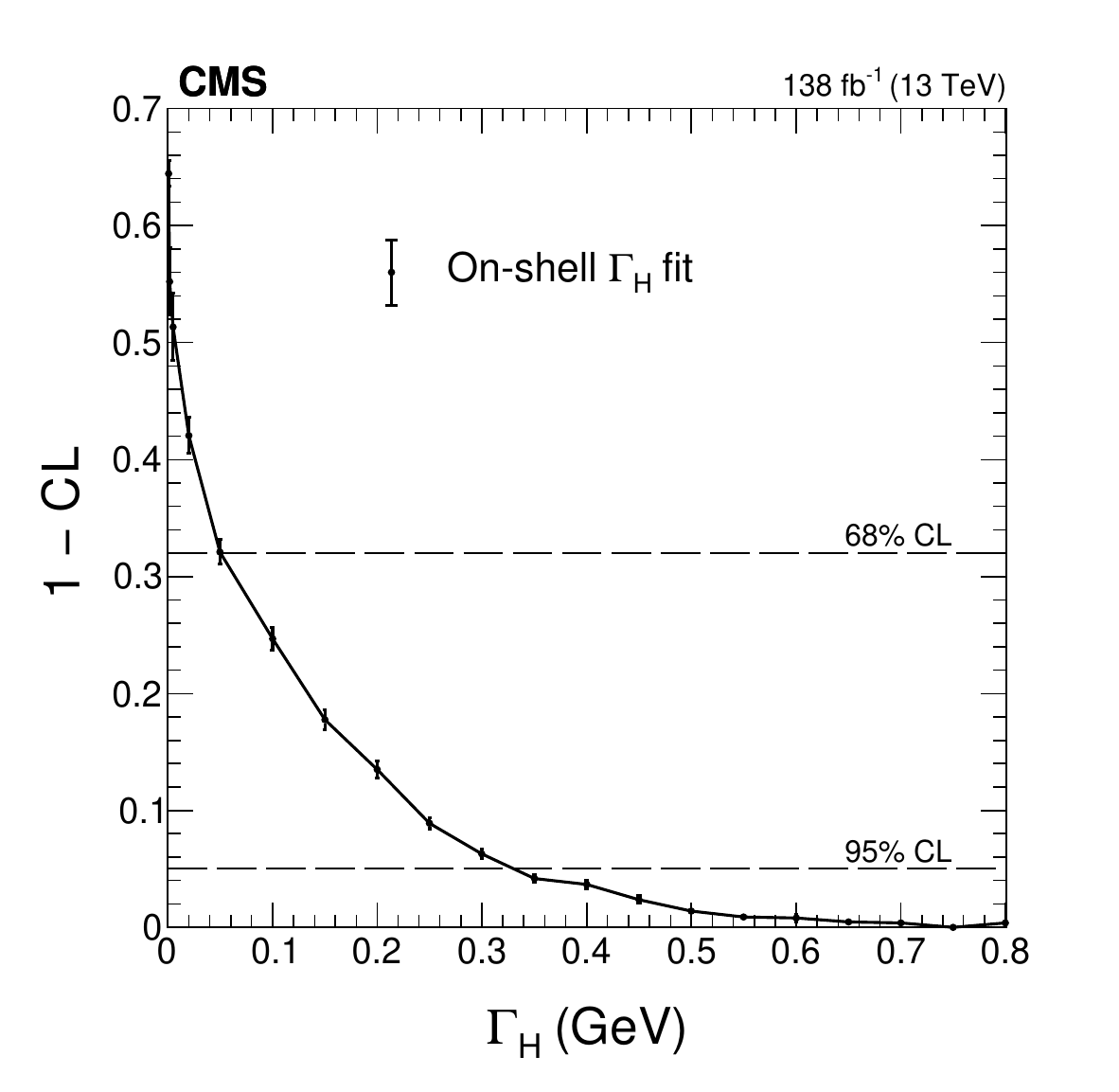}
  \caption{Distribution of $1-\CL$ vs. \GH from the fit in the measurement of the Higgs boson width using on-shell production only. The \CL values shown by the points are extracted using the Feldman-Cousins approach. The vertical bars on the points represent the spread of the simulated pseudo-experiment values. The 68 and 95\% \CL values are shown by the dashed horizontal lines.}
  \label{OnShell_width}
\end{figure}

The $\mathcal{N}$--2D$'_{BS}$ model is adopted also for the width measurement, changing the signal model, as described before, to include the \GH parameter. 
Since the theoretical value is very close to zero, which is a strict lower bound, confidence intervals for the \Hboson width are obtained following the Feldman--Cousins approach~\cite{FC}. 
The \CL is evaluated for several width hypotheses using distributions obtained from simulated pseudo-experiments. 
The observed (expected) upper limit on \GH is 50 (320)\MeV at 68\% \CL and 330 (640)\MeV at 95\% \CL. 
Although the observed limit is much less than the expected one, the two are statistically compatible.
The resulting distribution of $1-\CL$ vs. \GH is shown in Fig.~\ref{OnShell_width}.
The measurement precision is dominated by the statistical uncertainty and the subdominant systematic uncertainty is mainly driven by lepton momentum resolution.

\section{Higgs boson width measurement with off-shell production}

We perform an extended binned maximum likelihood fit to the on- and \offshell events split in several categories. 
The final measurements of \mH, \GH, and $\mu_{j}$ are conducted using the
 CMS statistical analysis tool \textsc{Combine}~\cite{CMS:2024onh}. 

The extended likelihood function is constructed using the probability densities in
Eqs.~(\ref{eq:ponshell}) and~(\ref{eq:poffshell}), with each event characterized by the discrete category $k$ and typically three 
observables $\vec{x}$. The likelihood $\mathcal{L}$ is maximized with respect to the nuisance parameters $\vec{\xi}_{jk}$ describing
the systematic uncertainties and the signal strength parameter $\mu$ (total signal strength), 
or \muF (signal strength for \ggH) and \muV (signal strength for the EW processes). The allowed 68 and 95\% \CL 
intervals are defined using the profile likelihood function, $-2\Delta\ln\mathcal{L} = 1.00$ and $3.84$, 
respectively, for which exact coverage is expected in the asymptotic limit~\cite{Wilks:1938dza}.

\begin{figure}[!htb]
  \centering
  \includegraphics[width=0.45\textwidth]{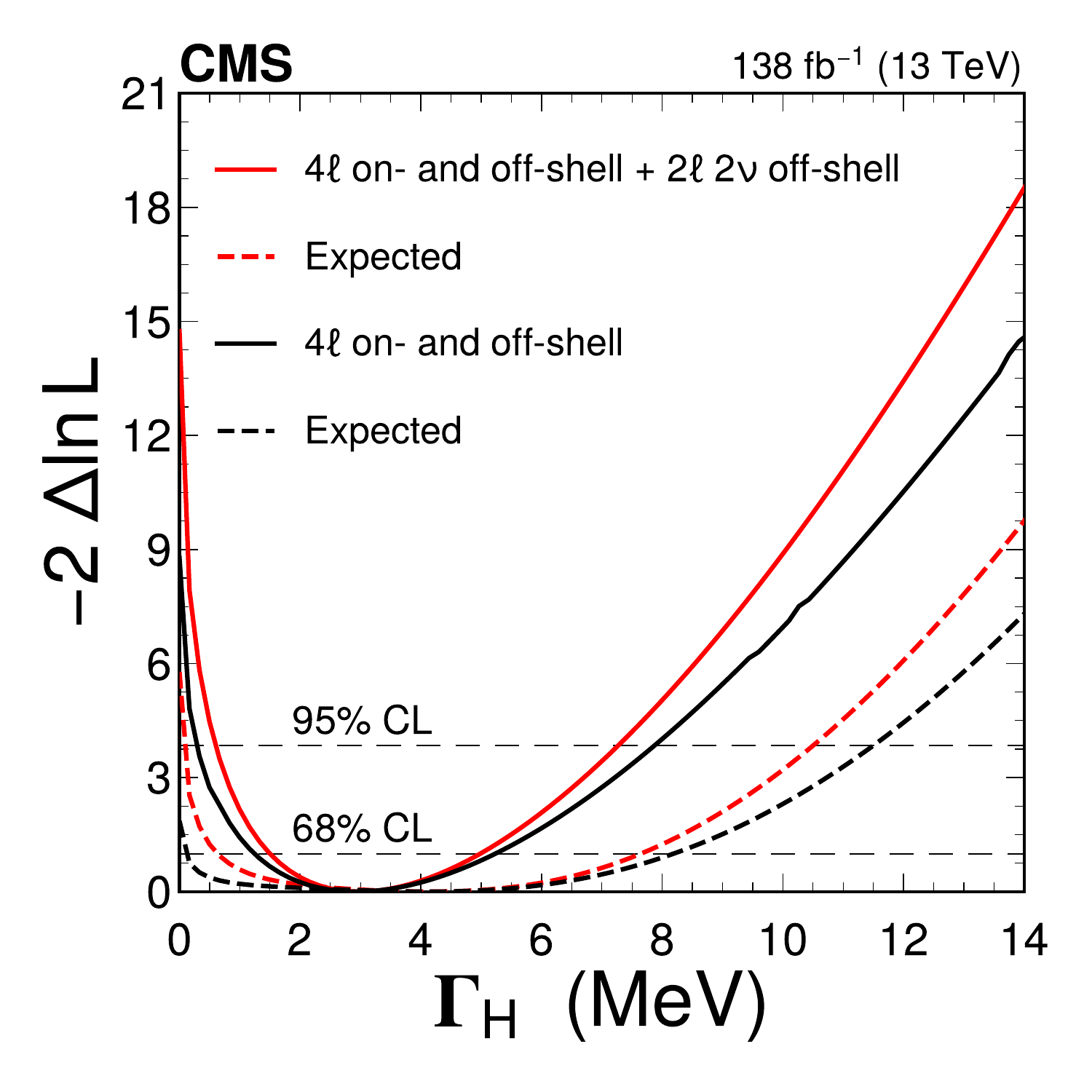}  
  \caption
      {
        Observed (solid) and expected (dashed) profile likelihood projections from the \Hboson width fit using on- and off-shell production from this analysis.
        The analysis of the \offshell $\PH\to\PZ\PZ\to4\Pell$ channel combined with the \onshell $\PH\to\PZ\PZ\to4\Pell$ channel~\cite{CMS:2021nnc} 
        is shown in black. The full combination of $\PH\to\PZ\PZ\to4\Pell$ with the \offshell $\PH\to\PZ\PZ\to2\Pell2\PGn$~\cite{CMS:2022ley} is given in red. 
        The black horizontal dashed lines show the 68 and 95\% \CL values. 
        \label{fig:widthscan} 
      }
\end{figure}

Constraints on \GH are set by simultaneously fitting $\PH\to\PZ\PZ\to4\Pell$ events from the on- and \offshell regions. 
The \onshell region corresponds to Scheme 2 in Ref.~\cite{CMS:2021nnc}, 
where six mutually exclusive event categories are defined and anomalous interactions are constrained to zero. 
It determines two signal strengths, $\mu_j$ in Eqs.~(\ref{eq:ponshell}) and~(\ref{eq:poffshell}), 
labeled as $\mu^\text{\onshell}_{\F}$ and $\mu^\text{\onshell}_{\PV}$, 
which correspond to production mechanisms driven by fermion and vector boson couplings of
the \Hboson, respectively, as defined in Section~\ref{sec:Introduction}. 
The \Hboson mass is constrained to $\mH=125.38$\,\GeV \cite{Sirunyan:2020xwk} in this fit. 
The observed and expected constraints on the \Hboson width are shown in Table~\ref{table:widthoffshellcomb}.
The likelihood scan of \GH using the asymptotic approximation method is shown in Fig.~\ref{fig:widthscan}. 
This measurement excludes the scenario of no \offshell \Hboson production with a \CL corresponding 
to 3.0 standard deviations (average expected 1.4).

\begin{table*}[!htb]
  \centering
  \topcaption{
    Summary of the total Higgs boson width \GH measurement, showing the 68\%~\CL (central values with uncertainties)
    and 95\%~\CL (in square brackets) intervals for the $\PH\to\PZ\PZ\to4\Pell$ channel alone and in combination with the 
    \offshell $\PH\to\PZ\PZ\to2\Pell2\PGn$ channel.
  }
  \renewcommand{\arraystretch}{1.25}
  \begin{scotch}{lll}
    Channel     & Observed \GH (\MeVns{})        &  Expected \GH (\MeVns{})  \\
    \hline
    $4\Pell$ on- and \offshell    & $2.9^{+2.3}_{-1.7} \ [0.3,7.9]$ & $4.1\pm 4.0 \ [<11.5 ]$ \\
    $4\Pell$ on- and \offshell  + $2\Pell2\PGn$ \offshell  & $3.0^{+ 2.0 }_{- 1.5 }  \ [0.6, 7.3]$ & $4.1\pm3.5 \ [0.1,10.5]$ \\
  \end{scotch}
  \label{table:widthoffshellcomb}
\end{table*}

The observed limits on \GH are stronger than the average expected values from simulation. 
This is supported by the upper left plot of Fig.~\ref{fig:ObservablesCat}, where the number of observed events in the 
sensitive region of $\mell>340$\GeV and $\Dbkg > 0.6$ in the Untagged category is below the expected value, but still consistent with it. 
The smaller number of events in this region favors the hypothesis of negative interference between the signal and background 
contributions, which dominates over the pure signal contributions for \GH values near the SM value. 
Therefore, large and very small values of \GH are disfavored.

\subsection{Examination of model dependency in width measurement}

The above \GH constraints assume the expected SM-like evolution of the \Hboson couplings over a large $m_{4\Pell}$ range.
The anomalous contributions to the $\PH{\PV\PV}$ vertex in EW production and \Hboson decay were evaluated in our earlier analyses 
using a smaller data set~\cite{Sirunyan:2019twz,CMS:2022ley}, and the constraints on \GH remained consistent.
However, the predominance of the top quark in the \ggH loop is assumed here and in our previous analyses. 
If there are additional contributions, such as from yet undiscovered heavy particles, then the $m_{4\Pell}$ evolution 
in the \offshell region would be altered.

To investigate the impact of large-mass yet undiscovered particles in the \ggH loop,
we introduce a new heavy quark~$\PQQ$ with an unconstrained coupling strength 
$\kappa_{\PQQ}$ in the likelihood parametrization, as described in Refs.~\cite{Gritsan:2020pib,Davis:2021tiv}.
In the framework of effective field theories, the contribution of~$\PQQ$ can be interpreted 
as a point-like interaction that encapsulates the influence of any heavy particles present in the loop.
The parameterization in Eq.~(\ref{eq:poffshell}) is extended to include the templates for terms proportional to 
$\kappa^2_{\PQQ}$ and $\kappa_{\PQQ}$, utilizing simulations reweighted with the \mela package
in the limit of the infinite $\PQQ$ mass. 
The $m_{4\Pell}$ shape in the \offshell region shows contrasting patterns between the SM \ggH production, 
which is mainly affected by the top quark loop with the $2m_{\PQt}$ threshold effect, 
and the \Hboson produced via \ggH loop involving the heavy quark~$\PQQ$~\cite{Gritsan:2020pib}.

An unconstrained $\kappa_{\PQQ}$ introduces additional uncertainty into the $m_{4\Pell}$ 
dependence in the \offshell region, resulting in less stringent limits on \GH.
However, both on- and \offshell $\PH\to4\Pell$ data constrain the possible values of $\kappa_{\PQQ}$, 
and the constraints on \GH remain largely consistent with those for $\kappa_{\PQQ}=0$.
The resulting \GH measurement from the combined on- and \offshell events is $2.7^{+2.7}_{-1.8} \MeV$ 
(expected $4.1^{+5.5}_{-4.1}\MeV$). 
The observed (expected) 95\%\, \CL interval is $\,[0.1,8.8]\, (\,[{<}14.4]\,)$\MeV.
We note that combining measurements from other \onshell \Hboson production and decay channels 
in the future will lead to much stricter constraints on $\kappa_{\PQQ}$, reducing the flexibility 
to alter the SM-like evolution of the \Hboson couplings over a large $m_{4\Pell}$ range.

\subsection{Higgs boson width measurement with a combination of off-shell channels}

The results of this analysis with the SM-like couplings are combined with the prior CMS \offshell $\PH\to\PZ\PZ\to2\Pell2\PGn$
analysis~\cite{CMS:2022ley}, giving the first CMS measurement of \GH using the full $4\Pell$ and $2\Pell2\PGn$ data sample 
collected during Run 2. 
The observed and expected \GH measurements are shown in Table~\ref{table:widthoffshellcomb} and Fig.~\ref{fig:widthscan}
and supersede the previous CMS results~\cite{CMS:2022ley} under the SM-like coupling assumption. 
The fit results rule out the scenario of no \offshell \Hboson production
with a \CL corresponding to 3.8 standard deviations (average expected 2.4).

\begin{figure}[!htb]
  \centering
  \includegraphics[width=0.48\textwidth]{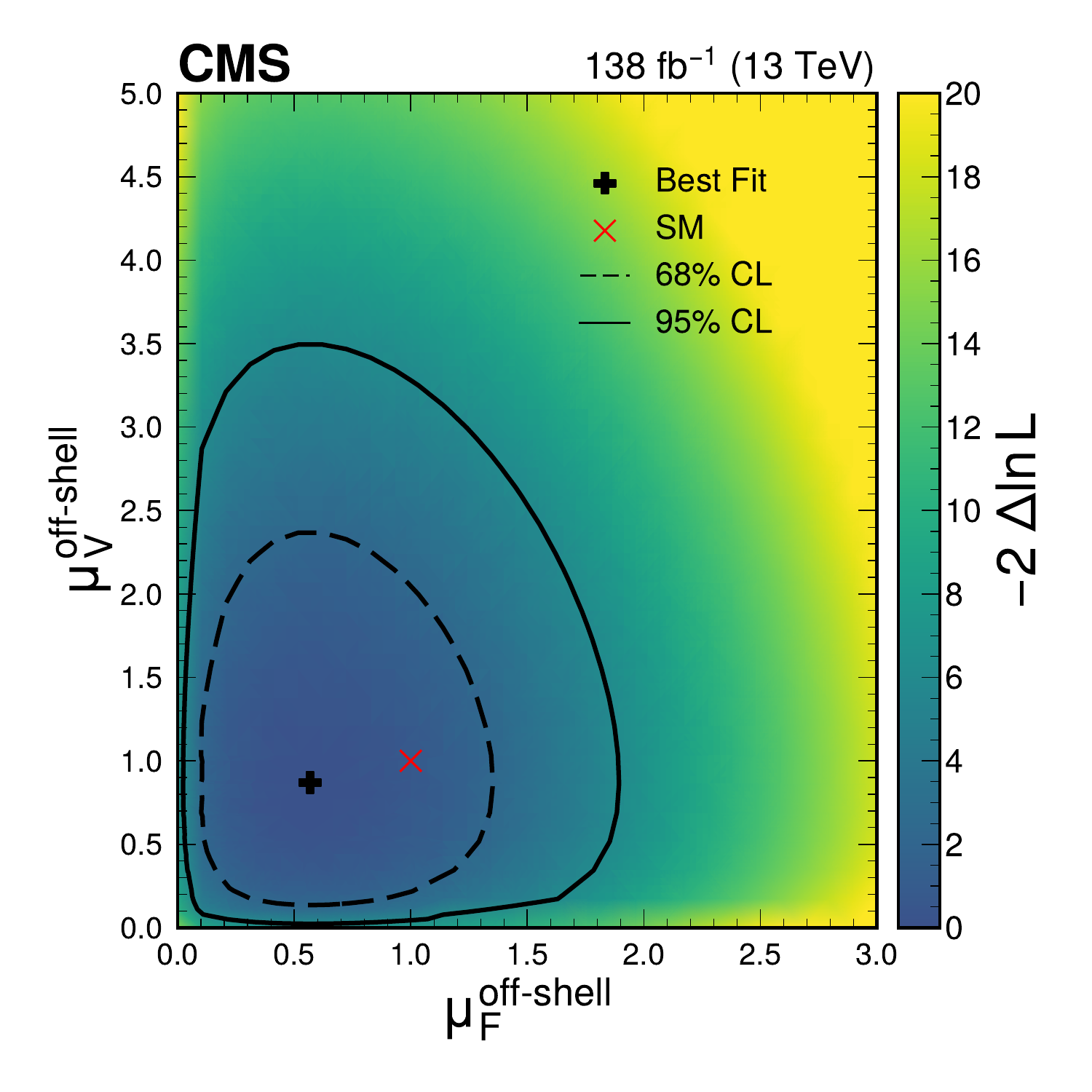}  
  \caption
      {
        Observed 2D profile likelihood projection of the \offshell signal strength parameters ($\mu^\text{\offshell}_{\F}$, $\mu^\text{\offshell}_{\PV}$) 
        from the fit to the combined \offshell $\PH\to\PZ\PZ\to4\Pell$ and $2\Pell2\PGn$ channels. The best fit value is shown by the black cross and the SM 
        prediction by the red x. The 68 and 95\% \CL contours are given by the dashed and solid curves, respectively. 
        The color scale to the right of the plot relates the quantitative values.
        \label{fig:muoffshell} 
      }
\end{figure}

The \offshell region fit can also be performed without relating its signal strength to that for the \onshell region.
In this case, the signal strength is modified by the parameter $\mu^\text{\offshell}$ common to all production 
mechanisms, with $\GH=\GHSM$ in Eq.~(\ref{eq:poffshell}), and the SM expectation corresponding to $\mu^\text{\offshell}=1$. 
In addition, we also perform a fit of the \offshell events with two unconstrained parameters $\mu^\text{\offshell}_{\F}$ and
$\mu^\text{\offshell}_{\PV}$, which express the signal strengths in the \ggH and EW processes, respectively.
The measured signal strengths are reported in Table~\ref{table:muoffshell} and a 2D scan of these parameters 
is presented in Fig.~\ref{fig:muoffshell}. The observed limits on signal strength in the \offshell region are stronger
than expected on average, following a trend similar to that previously discussed for \GH. 

\begin{table*}[!htb]
\centering
\topcaption{
  Measured values of the signal strengths $\mu^\text{\offshell}$, $\muF^\text{\offshell}$, and $\muV^\text{\offshell}$,
  and their 68\% and 95\% (in square brackets) \CL intervals from the combined fit to the \offshell $\PH\to\PZ\PZ\to4\Pell$ and $2\Pell2\PGn$ channels.
}
\renewcommand{\arraystretch}{1.6}
\begin{scotch}{lcc}
  Parameter                & {Observed}          &  {Expected}   \\
  \hline
  $\mu^\text{\offshell}$ &  $0.67^{+ 0.42}_{-0.32 }\ [0.14,1.54]$ & $1.00^{+ 0.83}_{-0.84}\ [0.02,2.46]$ \\
  $\muF^\text{\offshell}$ &  $0.57^{+ 0.50}_{- 0.36}\ [0.04,1.61]$ & $1.00^{+0.90}_{-0.98}\ [{<}2.62]$\\
  $\muV^\text{\offshell}$ &  $ 0.87^{+0.93}_{-0.57}\ [0.05,2.90]$ & $1.00^{+1.95}_{-0.89}\ [{<}4.34]$ \\
\end{scotch}
\label{table:muoffshell}
\end{table*}
                                                                                                                            
\section{Summary}
\label{Conclusion}
A measurement of the \Hboson mass (\mH) and width (\GH) using the decays to two \PZ bosons is presented.  The data sample comes from proton-proton collisions at the LHC recorded by the CMS experiment at a center-of-mass energy of 13\TeV, corresponding to an integrated luminosity of 138\fbinv. On-shell \Hboson production with the $\PH\to 4\Pell$ decay ($\Pell = \Pe$, $\PGm$) is used to measure its mass and constrain its width. The mass measurement yields $\mH=125.04 \pm 0.11\stat\pm0.05\syst\GeV = 125.04\pm0.12$\GeV, in agreement with the expected precision of $\pm$0.12\GeV. From \onshell production events, an upper limit of $\GH < 330\MeV$ is set at 95\% confidence level. The mass measurement is further improved by combining data from Runs 1 and 2, leading to the most precise single measurement of the mass to date in this channel,  $\mH=125.08\pm0.10\stat\pm0.05\syst\GeV = 125.08\pm0.12$\GeV. Using on- and \offshell \Hboson production with the decay to four leptons, and combining them  with a separate analysis with Higgs boson decay to two leptons plus two neutrinos, we measure  $\GH=3.0^{+2.0}_{-1.5}\MeV$, consistent with the standard model prediction of 4.1\MeV. These results are summarized in Table~\ref{table:widthoffshell}.  The strength of the \offshell Higgs boson production is also reported, and the scenario of no \offshell \Hboson production  is excluded at a confidence level corresponding to 3.8 standard deviations. Results of the measurements are tabulated in the HEPData record for this analysis~\cite{hepdata}.

\begin{table*}[!htb]
  \centering
  \topcaption{
    Summary of the Higgs boson mass and total width \GH measurements, showing the allowed 68\%~\CL (central values with uncertainties)
    and 95\%~\CL (in square brackets) intervals. Uncertainties are reported as a combination of statistical and systematic uncertainties. 
    The first two rows display the outcomes of the analysis conducted within the \onshell $\PH\to\PZ\PZ\to4\Pell$ region, 
    where the width is restricted to be positive. The third row incorporates results from the \offshell $\PH\to\PZ\PZ\to4\Pell$
    region combined with the \onshell $\PH\to\PZ\PZ\to4\Pell$~\cite{CMS:2021nnc} and \offshell $\PH\to\PZ\PZ\to2\Pell2\PGn$~\cite{CMS:2022ley}.
  }
  \renewcommand{\arraystretch}{1.6}
  \begin{scotch}{lcc}
    Parameter                & {Observed}          &  {Expected}   \\
    \hline
    \mH (\GeVns{})  &  $125.08 \pm 0.12$ & $\pm0.12$\\
    on-shell \GH (\MeVns{})  &  $<$50  $[{<}330]$  & $<$320 $[{<}640]$\\
    off-shell \GH  (\MeVns{})  & $3.0^{+2.0}_{-1.5}  \ [0.6, 7.3]$ & $4.1\pm3.5 \ [0.1,10.5]$ \\
  \end{scotch}
  \label{table:widthoffshell}
\end{table*}

\begin{acknowledgments}
We congratulate our colleagues in the CERN accelerator departments for the excellent performance of the LHC and thank the technical and administrative staffs at CERN and at other CMS institutes for their contributions to the success of the CMS effort. In addition, we gratefully acknowledge the computing centers and personnel of the Worldwide LHC Computing Grid and other centers for delivering so effectively the computing infrastructure essential to our analyses. Finally, we acknowledge the enduring support for the construction and operation of the LHC, the CMS detector, and the supporting computing infrastructure provided by the following funding agencies: SC (Armenia), BMBWF and FWF (Austria); FNRS and FWO (Belgium); CNPq, CAPES, FAPERJ, FAPERGS, and FAPESP (Brazil); MES and BNSF (Bulgaria); CERN; CAS, MoST, and NSFC (China); MINCIENCIAS (Colombia); MSES and CSF (Croatia); RIF (Cyprus); SENESCYT (Ecuador); ERC PRG, RVTT3 and MoER TK202 (Estonia); Academy of Finland, MEC, and HIP (Finland); CEA and CNRS/IN2P3 (France); SRNSF (Georgia); BMBF, DFG, and HGF (Germany); GSRI (Greece); NKFIH (Hungary); DAE and DST (India); IPM (Iran); SFI (Ireland); INFN (Italy); MSIP and NRF (Republic of Korea); MES (Latvia); LMTLT (Lithuania); MOE and UM (Malaysia); BUAP, CINVESTAV, CONACYT, LNS, SEP, and UASLP-FAI (Mexico); MOS (Montenegro); MBIE (New Zealand); PAEC (Pakistan); MES and NSC (Poland); FCT (Portugal); MESTD (Serbia); MCIN/AEI and PCTI (Spain); MOSTR (Sri Lanka); Swiss Funding Agencies (Switzerland); MST (Taipei); MHESI and NSTDA (Thailand); TUBITAK and TENMAK (Turkey); NASU (Ukraine); STFC (United Kingdom); DOE and NSF (USA).
        
\hyphenation{Rachada-pisek} Individuals have received support from the Marie-Curie program and the European Research Council and Horizon 2020 Grant, contract Nos.\ 675440, 724704, 752730, 758316, 765710, 824093, 101115353, 101002207, and COST Action CA16108 (European Union); the Leventis Foundation; the Alfred P.\ Sloan Foundation; the Alexander von Humboldt Foundation; the Science Committee, project no. 22rl-037 (Armenia); the Belgian Federal Science Policy Office; the Fonds pour la Formation \`a la Recherche dans l'Industrie et dans l'Agriculture (FRIA-Belgium); the F.R.S.-FNRS and FWO (Belgium) under the ``Excellence of Science -- EOS" -- be.h project n.\ 30820817; the Beijing Municipal Science \& Technology Commission, No. Z191100007219010 and Fundamental Research Funds for the Central Universities (China); the Ministry of Education, Youth and Sports (MEYS) of the Czech Republic; the Shota Rustaveli National Science Foundation, grant FR-22-985 (Georgia); the Deutsche Forschungsgemeinschaft (DFG), among others, under Germany's Excellence Strategy -- EXC 2121 ``Quantum Universe" -- 390833306, and under project number 400140256 - GRK2497; the Hellenic Foundation for Research and Innovation (HFRI), Project Number 2288 (Greece); the Hungarian Academy of Sciences, the New National Excellence Program - \'UNKP, the NKFIH research grants K 131991, K 133046, K 138136, K 143460, K 143477, K 146913, K 146914, K 147048, 2020-2.2.1-ED-2021-00181, and TKP2021-NKTA-64 (Hungary); the Council of Science and Industrial Research, India; ICSC -- National Research Center for High Performance Computing, Big Data and Quantum Computing and FAIR -- Future Artificial Intelligence Research, funded by the NextGenerationEU program (Italy); the Latvian Council of Science; the Ministry of Education and Science, project no. 2022/WK/14, and the National Science Center, contracts Opus 2021/41/B/ST2/01369 and 2021/43/B/ST2/01552 (Poland); the Funda\c{c}\~ao para a Ci\^encia e a Tecnologia, grant CEECIND/01334/2018 (Portugal); the National Priorities Research Program by Qatar National Research Fund; MCIN/AEI/10.13039/501100011033, ERDF ``a way of making Europe", and the Programa Estatal de Fomento de la Investigaci{\'o}n Cient{\'i}fica y T{\'e}cnica de Excelencia Mar\'{\i}a de Maeztu, grant MDM-2017-0765 and Programa Severo Ochoa del Principado de Asturias (Spain); the Chulalongkorn Academic into Its 2nd Century Project Advancement Project, and the National Science, Research and Innovation Fund via the Program Management Unit for Human Resources \& Institutional Development, Research and Innovation, grant B39G670016 (Thailand); the Kavli Foundation; the Nvidia Corporation; the SuperMicro Corporation; the Welch Foundation, contract C-1845; and the Weston Havens Foundation (USA).
\end{acknowledgments}

\bibliography{auto_generated}
\cleardoublepage \appendix\section{The CMS Collaboration \label{app:collab}}\begin{sloppypar}\hyphenpenalty=5000\widowpenalty=500\clubpenalty=5000
\cmsinstitute{Yerevan Physics Institute, Yerevan, Armenia}
{\tolerance=6000
A.~Hayrapetyan, A.~Tumasyan\cmsAuthorMark{1}\cmsorcid{0009-0000-0684-6742}
\par}
\cmsinstitute{Institut f\"{u}r Hochenergiephysik, Vienna, Austria}
{\tolerance=6000
W.~Adam\cmsorcid{0000-0001-9099-4341}, J.W.~Andrejkovic, T.~Bergauer\cmsorcid{0000-0002-5786-0293}, S.~Chatterjee\cmsorcid{0000-0003-2660-0349}, K.~Damanakis\cmsorcid{0000-0001-5389-2872}, M.~Dragicevic\cmsorcid{0000-0003-1967-6783}, P.S.~Hussain\cmsorcid{0000-0002-4825-5278}, M.~Jeitler\cmsAuthorMark{2}\cmsorcid{0000-0002-5141-9560}, N.~Krammer\cmsorcid{0000-0002-0548-0985}, A.~Li\cmsorcid{0000-0002-4547-116X}, D.~Liko\cmsorcid{0000-0002-3380-473X}, I.~Mikulec\cmsorcid{0000-0003-0385-2746}, J.~Schieck\cmsAuthorMark{2}\cmsorcid{0000-0002-1058-8093}, R.~Sch\"{o}fbeck\cmsorcid{0000-0002-2332-8784}, D.~Schwarz\cmsorcid{0000-0002-3821-7331}, M.~Sonawane\cmsorcid{0000-0003-0510-7010}, S.~Templ\cmsorcid{0000-0003-3137-5692}, W.~Waltenberger\cmsorcid{0000-0002-6215-7228}, C.-E.~Wulz\cmsAuthorMark{2}\cmsorcid{0000-0001-9226-5812}
% ...
\par}
\vskip\cmsinstskip
\dag:~Deceased\\
$^{1}$Also at Yerevan State University, Yerevan, Armenia\\
$^{2}$Also at TU Wien, Vienna, Austria\\
% ...
\end{sloppypar}
\end{document}